\documentclass[a4paper,11pt]{article}
\pdfoutput=1 

\usepackage{jheppub} 

\usepackage[T1]{fontenc} 

\usepackage{soul}

\def\lres{L_{\rm \small  {res}}}

\def\jraa{R_{\tiny{\rm AA}}^{\rm jet}}
\def\hraa{R_{\tiny{\rm AA}}^{\rm had}}

\def\pTo{{p_{\rm T,1}}}
\def\pTt{{p_{\rm T,2}}}

\def\aSC{{\kappa_{\rm sc}}}

\newcommand{\be}{\begin{equation}}
\newcommand{\ee}{\end{equation}}

\title{\boldmath Modification of Jet Substructure in Heavy Ion Collisions as a Probe of the Resolution Length of Quark-Gluon Plasma}

\author[a]{J. Casalderrey-Solana,}
\author[b,c]{G. Milhano,}
\author[d,e,f]{D. Pablos,}
\author[g]{and K. Rajagopal}

\affiliation[a]{Departament de F\'\i sica Qu\`antica i Astrof\'\i sica \& Institut de Ci\`encies del Cosmos (ICC), 
\\Universitat de Barcelona, Mart\'{\i}  i Franqu\`es 1, 08028 Barcelona, Spain}
\affiliation[b]{LIP, Av. Prof. Gama Pinto, 2, P-1649-003 Lisboa , Portugal}
\affiliation[c]{Instituto Superior T\'ecnico (IST), Universidade de Lisboa, Av. Rovisco Pais 1, 
\\1049-001, Lisbon, Portugal}
\affiliation[d]{Department of Physics and Astronomy, Wayne State University, Detroit MI 48201}
\affiliation[e]{Department of Physics and Astronomy, McGill University, Montr\'{e}al QC H3A-2T8}
\affiliation[f]{University of Bergen, Postboks 7803, 5020 Bergen, Norway}
\affiliation[g]{Center for Theoretical Physics, Massachusetts Institute of Technology, 
\\Cambridge, MA 02139 USA}

\emailAdd{jorge.casalderrey@ub.edu}
\emailAdd{gmilhano@lip.pt}
\emailAdd{Daniel.Pablos@uib.no}
\emailAdd{krishna@mit.edu}

\abstract{We present an analysis of the role that the quark-gluon plasma (QGP) resolution length, the minimal distance by which two nearby colored charges in a jet must be separated such that they engage with the plasma independently, plays in understanding the modification of jet substructure due to interaction with QGP.  The shorter the resolution length of QGP, the better its resolving power. We identify a set of observables that are sensitive to whether jets are quenched as if they are single energetic colored objects or whether the medium that quenches them has the ability to resolve the internal structure of the jet. Using the hybrid strong/weak coupling model, we find that although the ungroomed jet mass is not suitable for this purpose (because it is more sensitive to effects coming from particles reconstructed as a part of a jet that originate from the wake that the jet leaves in the plasma), groomed observables such as the number of Soft Drop splittings $n_{\rm SD}$, the momentum sharing fraction $z_g$, or the {\it groomed} jet mass are particularly well-suited to discriminate the degree to which the QGP medium resolves substructure within a jet.
In order to find the optimal grooming strategy, we explore different cuts in the Lund plane that allow for a clear identification of the regions of Soft Drop phase space  that enhance the differences in the jet substructure between jets in vacuum and quenched jets.
Comparison with present data seems to disfavor an ``infinite resolution length'', which is to say the hypothesis that the medium interacts with the jet as if it were a single energetic colored object. Our analysis indicates that as the precision of experimental measurements of jet substructure observables and the control over uncertainties in their calculation improves, it will become possible to use comparisons like this to constrain the value of the resolution length of QGP, in addition to seeing how the substructure of jets is modified via their passage through it.
}

\begin{document} 
\maketitle
\flushbottom

\section{Introduction}
\label{sec:intro}

QCD Jets are some of the most fascinating objects produced 
in high energy collisions.
These are sprays of particles 
produced as a consequence of the fragmentation of highly virtual partons produced in elementary collisions between energetic incident partons or electrons. 
While their microscopic origin as a manifestation of the 
structure of perturbative QCD has long been understood, it is only in recent years that a new suite 
of analysis techniques have opened the internal structure, in a generalized sense the shape, of jets
to quantitative scrutiny.
These new tools not only have allowed us to better characterize the radiation pattern of virtual quarks and gluons within jets
but also to design new observables with which to discover new physics \cite{Adams:2015hiv,Larkoski:2017jix}.

One of the areas where measurements of new observables that are sensitive to the internal shape and structure of jets shows the greatest promise is in the study of hot strongly interacting matter produced in ultrarelativistic heavy ion collisions.  The jets produced in such collisions
at the Relativistic Heavy Ion Collider and the Large Hadron Collider must form, shower, and propagate within the expanding, cooling, liquid droplet of hot quark-gluon plasma produced in the same
collisions. This means that the jets probe the hot medium, and as they do it means that
their own properties are modified.  The generic phrase for this phenomenon is jet quenching, but it should be kept in mind that the droplet of plasma is also modified as energy and momentum is exchanged between the jet and the droplet \cite{CasalderreySolana:2004qm,Neufeld:2008fi}.
The earliest such effect to be analyzed was the reduction in the
total energy of the jet.  However, much more can be learned about the interaction between
jets and the hot liquid, and ultimately about the properties of that liquid itself, by measuring and analyzing how the structure and shapes of the jets are modified via their passage through
the quark-gluon plasma.

An appealing property of jets in the study of dense hadronic matter is their potential to probe the short-distance-scale structure of the quark-gluon-plasma formed in those collisions. As multipartonic objects, jets can probe the in-medium physics at different transverse size, depending of the separation of its constituent partons as they traverse the medium.  
In studying the scattering of a single electron or parton off the medium, physics at
varying length scales can only be probed via analyzing scattering events with varying
momentum transfer.  
In the longer term, we can hope to use substructure observables to look for (uncommon) events
in which a single parton within a jet strikes a parton from the medium with sufficiently high
momentum transfer that a parton is scattered out of the jet at a large angle~\cite{DEramo:2012uzl,Kurkela:2014tla,DEramo:2018eoy}.  
These events
resolve the short-distance, partonic, structure of the QGP.  Before measurements of jet substructure
in heavy ion collisions can be used in this way, however, it is imperative to understand
how passage through the liquid QGP modifies the structure of jets, even without assuming that the microscopic partonic structure of the plasma is resolved.  
In this paper we take significant steps in this direction, studying how several different substructure observables
are modified  in a previously benchmarked model of jet quenching.

Furthermore, by analyzing
how entire jets interact with the medium as multipartonic objects, we introduce --- and have
a chance to take advantage of --- the length scales corresponding to the separations
between partons within a jet or between subjet constituents within a jet, as the jet propagates
through the medium.
Although we cannot yet use jets to resolve the short-distance structure of quark-gluon plasma, the notion of resolution arises in our analysis in quite a different way.  We can
ask whether the plasma is able to ``resolve'' the internal multipartonic structure of the jet itself or, better, to what degree it is able to do so.  
If a jet shower is narrow enough, we expect the medium to interact with it as if it were a single, unresolved, energetic probe, losing energy and momentum to the medium.
Only after the partons within the jet have separated sufficiently will they begin engaging
with the medium as independent objects.  The minimal distance by which
two nearby colored charges in a jet must be separated such that they engage
with the quark-gluon plasma independently is a property of the medium; we shall refer to
it as the resolution length of quark-gluon plasma and denote it by $\lres$.  
We expect $\lres$ to be parametrically of order $1/T$ if the medium is strongly coupled
or longer if it is weakly coupled, and thus not a microscopic aspect
of the structure of QGP. It nevertheless serves to characterize
the nature of this medium.

Our goals in this paper are thus twofold.  We seek to advance our understanding of how several jet substructure
observables are modified via the passage of the jets through the quark-gluon plasma
produced in heavy ion collisions.  And, we shall show how to take advantage of the length scales corresponding to 
the separation between subjet constituents within a jet to use experimental measurements
of groomed jet substructure observables to constrain $\lres$, a fundamental property 
of quark-gluon plasma.

In this paper we will provide a set of observables that clearly discriminate to what extent the inner structure of jets produced in heavy ion collisions at the LHC is modified by their interactions with the strongly coupled liquid produced in the same collisions. The main physics observation  that we will exploit in this paper is that as soon as the jet is resolved by the medium, and ``seen'' as more than one object that loses energy and momentum, the jet as a whole will
lose more energy. This happens because multiple components of the resolved jet each lose energy independently. 
This phenomenon has been 
inferred via comparison between experimental data and several rather different (perturbative~\cite{Milhano:2015mng}; strongly coupled~\cite{Rajagopal:2016uip,Brewer:2017fqy}; and hybrid~\cite{Casalderrey-Solana:2016jvj}) models for jet quenching.
However, in most  inclusive jet observables, such as the production rates for jets with a given $p_T$ or  correlations between jets or between a jet and a boson, the physics of how energy loss depends on
the number of resolved components in a jet, or the angular width of a jet, can be difficult
to isolate from other effects caused by the passage of the jet through the medium.
In this work, we will illustrate these difficulties by analyzing the probability distribution for the charged
 jet mass.
Among jets with a given energy, those which have a larger angular width have a larger jet mass.
Hence, we expect that jets with larger mass should lose more energy. Given
that more energetic jets are produced less frequently than less energetic jets, this means
that after passage through the plasma the jets with a given energy will be more likely to
be narrow, low mass, jets~\cite{Milhano:2015mng,Rajagopal:2016uip,Casalderrey-Solana:2016jvj,Brewer:2017fqy}.  In short, the jet mass distribution should shift toward lower masses. What we shall see in Section~\ref{sec:chmass}, though, is that there is a confounding
effect: the energy that the jet loses to the droplet of QGP creates a wake of moving liquid which, after hadronization, serves to widen the objects that are reconstructed as jets, making the jet
mass distribution shift toward larger masses.

After exploring this rather frustrating situation, we shall turn to jet substructure measurements
with which, we show, we can identify clear signatures of the consequences of 
the resolution of multiple components within a jet by the medium. 
We shall employ the `Soft Drop' procedure~\cite{Larkoski:2014wba}, 
a jet grooming technique designed to remove soft jet components 
most sensitive to the underlying event in proton-proton collisions
and that for us serves to greatly reduce the sensitivity to particles coming from the wake
in the medium. We shall explore two particular sets of groomed jet 
observables: the momentum sharing fraction distribution of two hard subjets within a jet 
and their invariant mass distribution. We show that the medium 
modification of these two observables differs significantly in a model in which 
the medium is able to resolve all partons within a jet ($\lres=0$) 
relative to that in a model in which the medium treats a jet as a single object ($\lres=\infty$).
According to our discussion above, when the jet constituents are resolved the groomed 
jet mass distribution shifts towards smaller values. In addition, we also find 
that resolved quenched jet samples possess a larger fraction of narrower jets 
and a smaller fraction of wider jets, also according to the general expectations above.  
As we shall explicitly test, grooming makes these two observables almost 
insensitive to the soft particles coming from the wake in the medium
that obscure the analogous effects in the charge jet mass distribution. This makes the groomed jet observables that we identify robust tools that can be used to discriminate
the power of the medium created in heavy ion collisions to resolve the structure
of multipartonic probes propagating through it, which is to say to constrain the 
value of $\lres$.

The theoretical analysis of jet-medium interactions has a long history (see Refs.~\cite{Mehtar-Tani:2013pia,Qin:2015srf} for recent reviews).  A substantial body of work has been devoted to understanding 
how high-energy partons and jets interact in a very-high-temperature perturbative QGP 
\cite{Baier:1996kr,Zakharov:1996fv,Baier:1998kq,Gyulassy:2000er,Wiedemann:2000za,Wang:2001ifa,Arnold:2002ja,Salgado:2003gb,Jeon:2003gi,Jacobs:2004qv,Lokhtin:2005px,CasalderreySolana:2007zz,Zapp:2008af,Zapp:2008gi,Lokhtin:2008xi,Armesto:2009fj,Schenke:2009gb,Majumder:2010qh,CasalderreySolana:2010eh,Wang:2013cia,Zapp:2013vla,Ghiglieri:2015zma,Blaizot:2015lma,Chien:2015hda,Cao:2017zih}.
However, at the temperatures reached by current colliders, the QGP is not weakly coupled and strong coupling effects become important to understanding in-medium jet dynamics. For this reason, our
analysis will be carried out within the framework of the hybrid strong/weak coupling model, developed in Refs.~\cite{Casalderrey-Solana:2014bpa,Casalderrey-Solana:2015vaa,Casalderrey-Solana:2016jvj,Hulcher:2017cpt,Casalderrey-Solana:2018wrw}. This is a phenomenological approach that seeks 
to separate the short distance dynamics of jets from 
the long distance physics of the QGP, which we take as strongly coupled. 
The model has been confronted with a large set of experimental data and it is able to describe and predict many inclusive jet observables. Another feature that will be instrumental for 
our purposes is that this is one of the few examples in the literature 
in which 
the ability of the medium to resolve multipartonic sources
are incorporated into a Monte Carlo event generator 
in the form of a single tunable parameter. As we will argue, while 
the concrete results that we will present in this work are predictions of this model, many of the systematics that we observe are generic and must be 
present in any jet-medium interaction framework that includes the resolution of  
jet constituents by the QGP, independent of the microscopic description of the interaction.

This paper is organized as follows. In Section~\ref{sec:model}, we briefly review the main features of the hybrid model that we employ throughout 
this paper. We will analyze  two extreme realizations of the model, one in which the medium can resolve all the partons within a jet, the fully resolved case where the medium has $\lres=0$, 
and one in which the jet-medium
interaction is independent of any internal structure within a jet and depends only
on its energy, the fully unresolved case where the medium has $\lres=\infty$.
In Section~\ref{sec:chmass}, we discuss the charged jet 
mass distribution and show that the competition between effects of quenching and 
effects originating from the backreaction of the jet on the medium 
makes it challenging to tease out the dependence of the modification
of this observable by jet-medium interactions 
on the resolving power of the medium
from measurements of this observable. In Section~\ref{sec:gob}, we focus on groomed jet observables. 
After describing the procedure and characterizing the number of Soft Drop splittings in 
Section~\ref{sec:nsd}, we analyze the momentum sharing fraction distribution in 
Section~\ref{sec:zgdist}, where we find that the absolutely normalized $z_g$-distribution 
can clearly discriminate between the two extreme limits. To further clarify this 
dependence and discuss new dynamics, in Section~\ref{sec:lundplane} we discuss the 
Lund plane defined by the first pair of subjets to pass the Soft Drop condition
and analyze the modification of this Lund plane distribution 
by media with different $\lres$. This analysis is used in Section~\ref{sec:gmass} to 
propose a new Soft Drop grooming procedure in which the groomed jet 
mass distribution is clearly sensitive to the dependence of quenching on jet 
substructure. To complete the analysis of groomed observables, in 
Section~\ref{sec:constrainingLres} we go beyond the two extreme assumptions and explore the sensitivity of the different observables to realistic values of the resolution length of the medium
 $\lres$. Finally, in Section~\ref{sec:conc} we compare our results to
 current measurements of groomed jet observables in heavy ion collisions and 
 compare our results to other analysis of these physics in the literature. 
 Several technical points are deferred to Appendices~\ref{sec:negatives} to \ref{sec:gmasspt}.

\section{The model}
\label{sec:model}

The hybrid strong/weak coupling model, introduced and extensively described in our earlier works Refs.~\cite{Casalderrey-Solana:2014bpa,Casalderrey-Solana:2015vaa,Casalderrey-Solana:2016jvj,Hulcher:2017cpt,Casalderrey-Solana:2018wrw}, addresses the challenge of describing the interaction, and consequent modification, of jets that traverse QGP with which they are concurrently produced in heavy-ion collisions. The model is based on the following key physical observations. 

First, that while the initial production of energetic partons and their showering occur at scales for which QCD is weakly coupled, both the physics of QGP and of its interaction with jet constituents involves scales of the order of the QGP temperature for which QCD is strongly coupled. In the hybrid model, the development of a jet while interacting with QGP is carried out by blending a holographic formulation of the strongly-coupled interaction of jet constituents with QGP into a perturbative QCD treatment of the production of hard partons and parton showering. A hard proton-proton collision
is generated with PYTHIA 8.23 \cite{Sjostrand:2014zea}, including initial state radiation but not multi-parton interactions,  showered down to a transverse momentum scale of 1 GeV, and kept at parton level. To take into account the different production rate of hard processes in PbPb collisions with respect to pp, initial state nuclear effects are accounted for by using modified parton distribution functions following the EPS09 parametrization~\cite{Eskola:2009uj}. The parton showers in these events are then endowed with a spacetime structure by assigning a lifetime to each individual parton according to \cite{CasalderreySolana:2011gx}
\begin{equation}
\label{eq:formtime}
	\tau = 2 \frac{E}{Q^2}\, ,
\end{equation}
where $E$ is the parton's energy and $Q$ its virtuality. 

The showered hard proton-proton event is then embedded into a background consisting of 
a simulated evolving, expanding and cooling droplet of QGP
obtained from state-of-art hydrodynamical simulations, averaged for a given centrality class~\cite{Shen:2014vra}. The origin of the hard event is determined 
probabilistically according to the density distribution of the nuclear 
overlap for the same centrality class, and its azimuthal orientation is set by sampling a uniform distribution in $[0, 2\pi)$. The  background provides local, that is for each space-time point, QGP temperatures necessary for the computation of the modifications imparted on 
parton showers by QGP. 

In our early works~\cite{Casalderrey-Solana:2014bpa, Casalderrey-Solana:2015vaa,Casalderrey-Solana:2016jvj}, each individual parton was taken to experience energy loss according to a rate computed holographically~\cite{Chesler:2014jva,Chesler:2015nqz} 
for light quarks in the strongly coupled plasma of ${\cal N} = 4$ supersymmetric 
Yang-Mills  (SYM) theory to be
\be
\label{CR_rate}
\left. \frac{d E}{dx}\right|_{\rm strongly~coupled}= - \frac{4}{\pi} E_{\rm in} \frac{x^2}{x_{\rm therm}^2} \frac{1}{\sqrt{x_{\rm therm}^2-x^2}}\,.
\ee
Here, 
\be
\label{CR_xtherm}
\quad \quad x_{\rm therm}= \frac{1}{2\aSC}\frac{E_{\rm in}^{1/3}}{T^{4/3}}
\ee
is the distance over which the initial energy $E_{\rm in}$ of a light quark is completely 
transferred to the QGP, assuming QGP with a constant temperature $T$.  In the hybrid model,
we take  a spatially and temporally varying
$T$ from the hydrodynamic background and apply (\ref{CR_rate}) to each parton,
with $T$ and hence $x_{\rm therm}$ varying as a function of $x$ along the parton trajectory.
The strength of interaction is controlled by the parameter $\aSC$ which depends on the 't Hooft coupling $g^2 N_c$, on details of the gauge theory, and on how the energetic parton is prepared. In the hybrid model, $\aSC$ is a free parameter to be fixed by comparing the model predictions with experimental data. 
This comparison is done at the hadronic level, after hadronizing the modified jet shower via the Lund string model as implemented in PYTHIA and without altering color flows between partons.
Fitted values of $\aSC$ result in a thermalization distance 3-4 times longer than in ${\cal N} = 4$ SYM  which is natural if considering the different number of degrees of freedom in QCD and ${\cal N} = 4$ SYM~\cite{Casalderrey-Solana:2014bpa,Casalderrey-Solana:2015vaa,Casalderrey-Solana:2018wrw}.  

Taking each individual parton in a shower to have its energy depleted according to Eq.~(\ref{CR_rate}) as in Refs.~\cite{Casalderrey-Solana:2014bpa, Casalderrey-Solana:2015vaa,Casalderrey-Solana:2016jvj} amounts to considering that the QGP 
resolves every parton in a shower from the very moment that each is produced in a splitting.
Conversely, applying Eq.~(\ref{CR_rate}) to a jet as a whole would
correspond to assuming that none of the jet internal structure is ever resolved by the QGP. 
Neither of these cases is realistic. Instead, the QGP should be able to resolve as 
separate objects with which it can interact independently only those partons which are 
sufficiently apart. 
This, the second key observation underlying the present implementation of the hybrid model was initially formulated in \cite{Hulcher:2017cpt} and was 
used in Ref.~\cite{Casalderrey-Solana:2018wrw} in demonstrating that experimental data on 
the suppression in the number of jets and of high-$p_T$ hadrons from heavy ion collisions at the LHC can all be fit simultaneously by adjusting $\aSC$. 
The resolution length $\lres$, which for a strongly coupled medium must be related to the inverse
of the local QGP temperature, sets the minimal transverse distance 
between partons in a shower such that they are seen as separate by QGP, meaning that
they lose energy independently.
Partons closer to each other than $\lres$ will interact with the QGP as a single object. 
Energy loss, as given by Eq.~(\ref{CR_rate}), is effected on each resolved shower component. 
The resolution length $\lres$ is a property of QGP that is just as fundamental as (and perhaps not very
different in magnitude than) its Debye screening length -- the minimal distance between static heavy partons 
such that they diffuse independently of each other in the QGP.
Note that in perturbative analyses of multiple partonic sources traversing the medium~\cite{MehtarTani:2010ma,MehtarTani:2011tz,CasalderreySolana:2011rz,CasalderreySolana:2012ef}, the resolution parameter can be related to the description of
how a single parton loses energy;
at strong coupling, no such connection between energy loss and the resolution length
has been established
and for this reason we shall take $\lres$ to be a second, independent, free parameter of our hybrid model that ultimately should be determined via comparison between calculations and experimental measurements of observables that are sensitive to its value. 

Our third, and last, key observation can be stated simply as that the energy deposited by jet constituents into the QGP is not necessarily lost from the jet reconstructed
from experimental data. In the hybrid model, the energy lost at the rate given by Eq.~(\ref{CR_rate}) for each resolved shower structure is, instantaneously and fully, thermalized -- subject to conservation of momentum as well as energy. In other words, 
as the partons in the jet pass through the expanding liquid they locally heat and boost some of it, leaving behind a wake in the QGP. 
Overall conservation of energy-momentum requires that the hydrodynamic response of the plasma, the wake of a jet, to the the energy transferred by the jet carries momentum in the direction of motion of the jet.
While this contribution has been implemented in several models in different approximations \cite{He:2015pra,Chen:2017zte,Tachibana:2017syd,He:2018xjv,Park:2018acg,Chang:2019sae}, in the hybrid model  energy-momentum conservation is implemented in a simplified way without introducing additional parameters. 
Noting that the total energy deposited by a jet onto the QGP is small when compared with the total energy per unit rapidity in the event, we take the additional momentum acquired by the plasma as a small perturbation which, in turn, results in small perturbations to the final particle distribution. 
The modification to the particle spectrum due to the passage of a jet, computed and fully described in Ref.~\cite{Casalderrey-Solana:2016jvj} upon making these assumptions, is given by
\begin{equation}
\label{onebody}
\begin{split}
E\frac{d\Delta N}{d^3p}=&\frac{1}{32 \pi} \, \frac{m_T}{T^5} \, \rm{cosh}(y-y_j)  \exp\left[-\frac{m_T}{T}\rm{cosh}(y-y_j)\right] \\
 &\times \Bigg\{ p_{\perp} \Delta P_{\perp} \cos (\phi-\phi_j) +\frac{1}{3}m_T \, \Delta M_T \, \rm{cosh}(y-y_j) \Bigg\}.
\end{split}
\end{equation}
where $p_T$, $m_T$, $\phi$ and $y$ are the transverse momentum, transverse mass, azimuthal angle and rapidity of the emitted thermal particles whose distribution we have obtained,
and where $\Delta P_T$ and $\Delta M_T=\Delta E/\cosh y_j$ are 
the transverse momentum and transverse mass transferred from the jet (whose azimuthal angle and rapidity
are  $\phi_j$ and $y_j$) to the wake in the fluid.   
Comparisons between the calculations  in Ref.~\cite{Casalderrey-Solana:2016jvj}
and experimental measurements of observables including the jet shape and missing-$p_T$ observables
indicate that Eq.~(\ref{onebody}) predicts a few too many very soft particles coming from the wake and not quite enough such particles in the 2-4 GeV range of $p_T$; this may indicate
that the wakes left behind in droplets of QGP by passing jets do not fully thermalize.

A characteristic feature of our expression~(\ref{onebody}) for the change in the distribution of particles created in the freezeout of the QGP-fluid after the jet passage 
is that it can become negative. This is not unphysical. This negative contribution reflects the fact that when a jet deposits momentum in the plasma, the fluid is pushed in the direction of that jet. As a consequence of this overall motion of the fluid, when the back-reacted fluid freezes out, it emits more particles along the jet direction than the average event. Conversely, the production of particles in the direction opposite to the jet is reduced. Therefore, if we use the average event to separate the jet and medium particles in any observable, we will over-subtract some particles that 
are correlated with the jet.

Eq.~(\ref{onebody}) yields the average number of particles produced by the fluid in association with the jet. For our jet analyses, in which jets are reconstructed via sequential algorithms, we generate medium particles according to the distribution via a Metropolis algorithm described in Ref.~\cite{Casalderrey-Solana:2016jvj}. 
To take into account the over-subtraction effect described above, the particles coming from Eq.~(\ref{onebody}) that are associated with negative regions of the modified spectrum are considered as carrying negative mass and momentum. This mimics the effect of over-subtracting particles correlated with the jet. The technical details of how these negative particle contribution are treated in our jet analyses can be found in Appendix \ref{sec:negatives}. 

\subsection{Fixing the parameters of the model}

As we have just described, our implementation of the model into a Monte Carlo simulation depends on two model parameters: the strength of the energy loss, parametrized by $\aSC$, and the medium resolution scale $\lres$. Since an important part of this work will be to study how different observables can be used to shed light on the question of whether the medium resolves
components of a jet, we 
will explore our model for two extreme values of $\lres$: $\lres=0$ which represents the case in which all partons in the jet are resolved by the medium right from the moment when they are produced in a splitting; and $\lres=\infty$ which is the limit in which the resolution length is so large that none of the partons that form the jet is separately resolved and throughout the evolution and propagation of the jet it is seen by the medium as a single object. 
Neither of these choices is realistic.
We will also show results for the value $\lres=2/(\pi T)$, which lies within the range of plausible values for this parameter~\cite{Hulcher:2017cpt}.
For each of these different realizations of our model, we shall use different values of the 
quenching parameter $\aSC$, fitted to data as we describe below. 

The case where all partons in the jet are separately resolved from the moment that they are produced in a splitting, namely $\lres=0$, corresponds to the choice that we made in much of our early work with the hybrid model~\cite{Casalderrey-Solana:2014bpa,Casalderrey-Solana:2015vaa,Casalderrey-Solana:2016jvj}.  For this realization of the model, we have a reliable determination of the value of $\aSC$, obtained via a global fit to experimental data on the suppression in the number of jets and high-$p_T$ hadrons in Ref.~\cite{Casalderrey-Solana:2018wrw}.  
The simultaneous description of all those data fixes $0.404< \aSC< 0.423$, in the case where we assume that parton energy loss turns off below a temperature
of $T_c=145~$MeV. 
As explained in
 Ref.~\cite{Casalderrey-Solana:2018wrw}, the fact that the QCD transition is a continuous crossover implies that there is not a well defined value of $T_c$. 
 However, an equivalently good fit can be extracted by setting $T_c=170~$MeV, yielding a slightly larger range for the model parameter of $0.447<\aSC<0.470$~\cite{Casalderrey-Solana:2018wrw}. 
 A similar fitting procedure was performed for $\lres=2/(\pi T)$ which for $T_c=145~$MeV  ($T_c=170$) MeV yields $0.428<\aSC<0.447$ ($0.472<\aSC<0.494$).  
 Since after refitting the overall results are equivalent, all the results of this paper are presented for a fixed value  $T_c=145~$MeV. 
 
We have not performed a global fit analysis for the other extreme case in which
$\lres=\infty$ and the medium sees each jet as a single object throughout its showering and propagation. This is because in  this limit 
the predictions of the model for $\hraa$ and $\jraa$ are much closer to one another than is observed in experiments, 
as we show in Fig.~\ref{Fig:raa} in Appendix~\ref{sec:raa}.
Therefore, it is not possible to simultaneously describe the LHC data sets on hadron suppression and jet suppression with our model with $\lres=\infty$.  
This can already be taken as an indication that at least some aspects of the substructure of jets are resolved by the medium.  Our goal in this paper is to obtain more direct evidence for this conclusion, however.
To make comparisons of predictions for substructure observables in the case where $\lres=\infty$ to those where $\lres=0$, we shall choose $\aSC$ in the former case such that the jet suppression is the same as that in the latter case for jets with $p_T=150$~GeV.  This choice corresponds to choosing
 $0.5< \aSC< 0.52$  when we work with $\lres=\infty$. Details of how we have fixed these values can be found in Appendix~\ref{sec:raa}.

With these specifications, all parameters of the model are fixed and we can study  different jet observables using analysis strategies similar to those used in experiments.

\section{Charged jet mass}
\label{sec:chmass}

As we have stressed in the Introduction, the total energy lost by a jet in the medium can depend on the jet substructure, as long as this jet 
shower develops during its passage through the medium. As observed in many dynamical realizations of jet quenching \cite{Milhano:2015mng,Rajagopal:2016uip,Casalderrey-Solana:2016jvj,Brewer:2017fqy}, wide jets with 
multiple fragments lose more energy per distance traveled than narrow jets with the same energy that have fewer internal components,  since the wide jets have more sources of energy loss. 
The magnitude of this difference is controlled by the ability of the medium to resolve the internal structure of the jet \cite{Hulcher:2017cpt, Mehtar-Tani:2017web, Milhano:2015mng}. As already mentioned, in our model this ability is controlled by the medium resolution parameter $\lres$.

A consequence of the above physics reasoning is that
in the realization of our model with perfect resolution
($\lres=0$) the ensemble of jets after quenching will on average be narrower than
in the realization of our model in which jet substructure is never resolved
($\lres=\infty$). This is because, if their substructure is resolved, wider jets possess more fragments
traversing the medium than narrower ones, and are therefore more quenched.   In contrast,
if $\lres=\infty$ and jets are never resolved then their quenching is independent of their substructure and there is no preferential quenching of wider jets.
Since the wider jets have a larger jet mass, a natural expectation is that as a consequence of this physics
the mass distribution of the ensemble of jets after quenching will be shifted toward lower masses in the realization of our model with perfect resolution than in the realization in which jet substructure is never resolved.  In the latter case, we should expected a mass distribution after quenching which is closer to the mass distribution of jets in vacuum, without any quenching.
As we will see in this Section, soft dynamics associated with the wake
that the jets leave behind in the droplet of QGP
obscures this effect. 

To be able to compare with ALICE results  from Ref.~\cite{Acharya:2017goa}, we compute the charged jet mass, defined as 
\be
M^2_{\rm ch} \equiv (p^{\mu}_{\rm jet})^2 \ ,
\ee 
in our model.  Here, jets are reconstructed with the anti-$k_T$ algorithm~\cite{Cacciari:2008gp} with radius parameter $R=0.4$ using only charged hadrons that lie within $|\eta|<0.9$, and the jet is required to be within $|\eta|<0.5$.

In the experimental analysis~\cite{Acharya:2017goa}, a constitutive subtraction method is applied to the momentum of each of the jet particles to eliminate the contribution of the many soft particles produced in PbPb events. Unlike in this experimental analysis, in our simulations we do not perform such a subtraction since we can tag which particles are correlated with the jet and restrict our analysis to those.  In our model these can be broadly classified into two sets: particles that arise from the hadronization of the quenched jet shower; and particles that arise from the moving fluid left behind by the  back-reaction of the jet on the fluid. Unlike the former, the latter are 
generically very soft, since they have a transverse momentum comparable to the medium temperature; nevertheless, as we will see these have a significant effect on this observable.

\begin{figure}
\centering 
\includegraphics[width=1\textwidth]{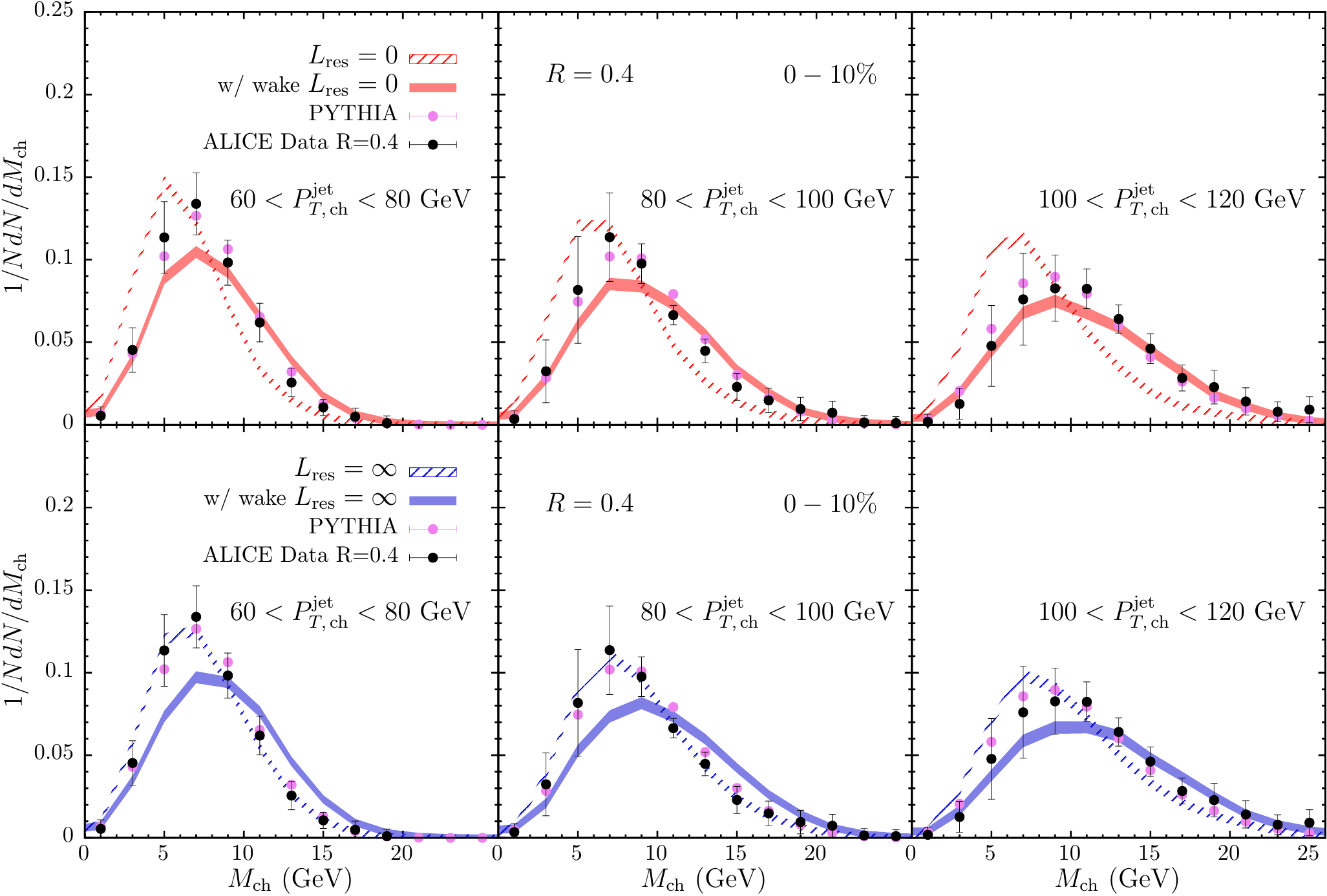}
\caption{\label{Fig:chmass} Results for the charged jet mass from hybrid model computations at $\sqrt{s}=2.76$ ATeV, for two extreme values of $\lres$, with $\lres=0$ in the upper row and $\lres=\infty$ in the lower row, compared to ALICE unfolded PbPb data \cite{Acharya:2017goa}. Results for our vacuum reference, labeled as PYTHIA, are shown in purple dots. Dashed bands correspond to our results without the inclusion of particles coming from medium response, while solid bands correspond to the full result.}
\end{figure}

In Fig.~\ref{Fig:chmass} we show our hybrid model results for the distribution of the charged jet mass before quenching (PYTHIA in the figure) and after quenching, for two values of $\lres$ (the extreme possibilities $\lres=0$ and $\lres=\infty$), comparing these results to measurements of PbPb collisions from ALICE. In order to focus solely on the effects of resolution, we start by discussing 
the contribution of particles from the jet hadronization, without including the soft medium back reaction contribution.
This is represented by the dashed lines in all panels of Fig.~\ref{Fig:chmass}. Consistent with the expectation described above, 
for $\lres=0$  we observe that charged jet mass distribution clearly shifts to the left, consistent with a narrowing of the final jet distribution for the given jet $p_T$ range.  In contrast, the choice 
of $\lres=\infty$ translates into barely any modification of the charged jet mass distribution, which is again consistent with our discussion. 
When comparing with ALICE data, which show that the PbPb mass distribution is consistent with the proton-proton distribution as simulated by PYTHIA we may be tempted to conclude from this data that the resolution scale in the plasma formed in PbPb collisions must be large. However, reaching such a conclusion would be premature.

As shown in the solid bands of Fig.~\ref{Fig:chmass}, the sensitivity of this observable to the soft particles that originate from the hadronization of the moving fluid that carries the momentum
lost by the jet is significant.
Let us stress here that including this set of particles correlated with the jet direction is not optional.
Momentum conservation makes it a necessity that when the jet loses momentum there must be a region of the droplet of plasma that gains this momentum, and it is then a necessity that, after this moving fluid hadronizes, some fraction of the hadrons which result end up reconstructed as a part of what the experimentalists define as a jet.
The inclusion of  these  soft particles, which are distributed almost uniformly in angle 
out to some angle that is much larger than the anti-$k_T$ jet reconstruction parameter $R=0.4$,
pushes the jet mass distribution out toward larger masses, hence acting in the opposite direction
to the effect that we discerned above.  Recall that the previous effect is much larger for $\lres=0$
than for $\lres=\infty$; unfortunately, the push to larger jet masses coming from the particles originating from the wake in the plasma is also larger for $\lres=0$, for the simple reason that the energy loss experienced by the wider jets -- and hence the wake that they leave in the plasma -- is greater when $\lres=0$.
The competition between these two opposite effects 
conspires to yield the conclusion that in our hybrid model with $\lres=0$ and the back-reaction
on the medium (the wake in the plasma) included the jet mass distribution after quenching is consistent within uncertainties with what it was for jets in vacuum, absent any quenching.
This in turn is consistent with ALICE unfolded data for jets in PbPb collisions
across the three different ranges of jet $p_T$ shown in Fig.~\ref{Fig:chmass}.  
In contrast, in the model realization in which $\lres=\infty$ and the substructure of jets is
never resolved by the medium the amount of energy lost has little sensitivity to the jet mass,
and neither does the strength of the wake in the plasma.
With $\lres=\infty$ it turns out on balance that, upon including the particles coming from the wake in the
plasma, the jet mass distribution after quenching is somewhat wider than that observed in the 
data\footnote{As may be inferred by inspection of Fig.~\ref{Fig:chmass}, the normalization of the  jet mass distribution after quenching differs between the calculations that include or omit the soft particles coming from the wake in the plasma. The reason for this is the presence of ``negative particles'' associated with back-reaction, which mimic the effect of over-subtraction of a homogeneous background~\cite{Casalderrey-Solana:2016jvj}, and which in some cases (about 5\% of the sample) lead to negative squared masses. 
While these ``imaginary'' masses cannot be shown in a standard plot, 
they are responsible for the fact that the curves in Fig.~\ref{Fig:chmass} integrate to somewhat less
than one.
}.

While the agreement 
of our hybrid model with this data in the totally resolved limit with $\lres=0$ seems at first glance rather striking, 
it is important to realize that this charged mass distribution arises as the direct consequence of the combination of two competing effects that, within uncertainties, cancel in our model.  
On the one hand, we expect that when the substructure of jets are resolved the ensemble of jets after quenching should be narrower, with a jet mass distribution pushed to lower masses; on the other hand, the medium back reaction introduces many soft particles spread over a large angle which pushes the jet mass distribution upwards. 
Noting that in previous work we have shown quantitatively that our model treatment of the medium backreaction
is incomplete in ways such that our model fails to describe measurements of the modification
of the jet shape in PbPb collisions~\cite{Chatrchyan:2013kwa}, it is appropriate to suspect that the cancellation that we have found here is only an artifact of the model.
 The competition between two effects pushing in opposite directions, and their near cancellation, suggests that the jet mass distribution is not an observable that is well suited to determining
 whether the medium produced in heavy ion collisions does or does not resolve the substructure of 
 jets shooting through it.  We need to find other observables that have significantly less sensitivity
 to the soft particles coming from the hadronization of the flowing wake that the jets leave behind in the
 plasma, since such observables should in general be under better theoretical control.
With this as motivation, in the 
 next Section we will concentrate on groomed jet substructure observables that are dominated by hard components of a jet.

\section{\label{sec:gob}Groomed jet observables}

Over the last decade, a large number of jet analysis techniques have been developed in order to devise observables with decreased sensitivity to soft components of jets. 
These techniques, generically known as jet grooming methods, have been used extensively to analyze jets in proton-proton collisions. In the last few years, some of those tools --- in particular those employing the `Soft Drop' procedure~\cite{Larkoski:2014wba} --- have also been employed to study jets produced in heavy ion collisions, both at the LHC~\cite{Sirunyan:2017bsd,Sirunyan:2018gct,Acharya:2019djg} and at RHIC~\cite{Kauder:2017cvz}.

Motivated by these measurements, in this Section we will explore the sensitivity of different observables obtained via `Soft Drop' to choices we make for the resolution length of QGP. 
Althought the algorithm and its properties have been thoroughly described in the literature \cite{Larkoski:2014wba}, we shall choose here to describe the procedure as implemented in our Monte Carlo simulations both for completeness and to establish notation. 

The Soft Drop procedure starts by identifying and reconstructing a sample of jets using the anti-$k_T$ algorithm~\cite{Cacciari:2008gp} with some specified value of the anti-$k_T$ radius $R$.
The constituents of each jet that have been identified in this fashion are then reclustered using
the Cambridge/Aachen algorithm~\cite{Dokshitzer:1997in,Wobisch:1998wt}.
This algorithm, which is based only on the angular separation of tracks not on their energies or momenta,  is not suitable for finding jets; speaking colloquially, the reason for this is that if it is used to find jets this 
algorithm gets much too easily distracted by soft particles from the underlying event (in pp collisions) and the medium (in heavy ion collisions).
That said, however, once a jet has been identified, reclustering its constituents using the Cambridge/Aachen algorithm is particularly effective at
finding its subjets.    
In particular, for an angular ordered shower as in a jet that showers in vacuum, in which each subsequent splitting happens to leading logarithmic accuracy at a smaller angle than that of the splittings that preceded it,
the Cambridge/Aachen algorithm in effect seeks to rebuild the jet in the opposite order from the way in which it developed in time.  The reclustering first groups constituents that are closest together, as for those originating from the last splittings in the shower.  It then clusters these groups sequentially into structures that are larger and larger in angular extent, with the last steps in the reclustering thus plausibly corresponding to the earliest splittings in the parton shower unless, that is, they originate from soft gluon radiation.
With this motivation in mind, the next step in the Soft Drop procedure is to separate the jet into two subjets by undoing the last step of the Cambridge/Aachen reclustering, and then to 
check whether the configuration of two subjets satisfies a `Soft Drop condition'.  If the condition is not satisfied, this means that one of the two subjets is considered soft, and is dropped.  The procedure
is then repeated until the Soft Drop condition is satisfied, at which point the two subjets then in hand are used to defined groomed jet observables.
Different variants of the Soft Drop procedure, corresponding to different choices for the Soft Drop condition and hence different criteria for what is groomed away, can be employed.
In particular, the configuration of two subjets is said to pass the `Soft Drop condition'
if  these two subjets satisfy
\be\label{SoftDropCondition}
\frac{{\rm min}  \left(\pTo,\, \pTt\right)}{\pTo+\pTt} > z_{\rm cut} \left(\frac{\Delta R}{R}\right)^\beta
\ee
with $\Delta R$ the angular separation between the subjets in the $\left(\eta\,,  \phi \right)$ plane and $z_{\rm cut}$ and $\beta$ two parameters that specify 
the grooming procedure,
controlling which emissions are groomed away. 
If the pair of subjets does not satisfy the condition, then the softer subjet is dropped and the process is continued by undoing one more step of the Cambridge/Aachen reclustering of the harder branch, and repeating the check of the Soft Drop condition again.
If at some point (after some number of grooming stages) a configuration of two subjets is found
that satisfies the condition (\ref{SoftDropCondition}), this means that the jet contains two subjets neither of which is soft, and these two subjets may then be used to defined groomed jet observables.
If, however, as one undoes all the steps of the Cambridge/Aachen reclustering  
the Soft Drop condition~(\ref{SoftDropCondition}) is never met, the jet is said to be untagged, which in this context means that it cannot be separated into two subjets neither of which is soft. 

Of the two parameters in the Soft Drop condition (\ref{SoftDropCondition}), the one whose role is more straightforward to understand is $z_{\rm cut}$.  If a very small value of $z_{\rm cut}$ is chosen, almost anything satisfies the Soft Drop condition and almost nothing is groomed away.
Choosing a sufficiently large $z_{\rm cut}$ 
ensures that the momenta of two subjets satisfying (\ref{SoftDropCondition}) that remain after
grooming
are both large relative to momenta typical of the
underlying event, meaning that the properties of these two subjets become less and 
less sensitive to soft dynamics, and in the case of a heavy ion collision less sensitive to particles originating from the hadronization of the droplet of plasma rather than from  the jet.  The groomed observables become more and more insensitive to soft dynamics for larger and larger values of $z_{\rm cut}$ and, for any given $z_{\rm cut}$, for jets with higher and higher jet energy.

Choosing the parameter $\beta$ gives us a means for intoducing a preference for smaller or larger angles between the pairs of subjets that remain after grooming.  If $\beta=0$~\cite{Dasgupta:2013ihk} then soft fragments are groomed away without regard for their angular separation.  If $\beta<0$ then slightly harder fragments will get groomed away if they are more collinear while slightly softer fragments that
are separated by a larger angle may satisfy the Soft Drop condition (\ref{SoftDropCondition}) .
If $\beta>0$, soft but nearly collinear subjets can satisfy the Soft Drop condition.
From this discussion, it becomes clear that only choices with  $\beta < 0$ are collinear safe. Other values of $\beta$ yield groomed observables that are nevertheless Sudakov safe \cite{Larkoski:2015lea}  and will also prove instructive at various points below.

If the Soft Drop procedure yields two groomed subjets satisfying the condition~(\ref{SoftDropCondition}), the first observable that we define
is the momentum-sharing fraction of those two subjets, denoted $z_g$ and defined as the right-hand side of the condition~(\ref{SoftDropCondition}):
\be\label{zgdef}
z_g \equiv \frac{{\rm min}  \left(\pTo,\, \pTt\right)}{\pTo+\pTt} \, .
\ee
By construction the maximum value for this fraction is $z_g=0.5$. Its lowest value depends on the parameter $\beta$ and it is $z=z_{\rm cut}$ for $\beta=0$. 

From the kinematics of the two groomed subjets that first satisfy the Soft Drop condition (\ref{SoftDropCondition}), we can also define a Soft Drop groomed jet mass $M_g$ as follows.  We first define the groomed four-momentum of the jet:
\be
\label{Pgdef}
P_g\equiv P_1 + P_2\ ,
\ee
where $P_1$ and $P_2$ are the four momenta of the subjets with
transverse momenta $\pTo$ and $\pTt$.
The groomed mass $M_g$ is then defined in a straightforward fashion
as the corresponding invariant mass: 
\be
\label{gmdef}
M_g^2\equiv P_g^2\, .
\ee
We shall present results for the distributions of $z_g$ and $M_g$ from our model calculations done with varying $\lres$ in Sections~\ref{sec:zgdist} and~\ref{sec:gmass} respectively.

We shall find that the modification of the internal structure of jets that propagate through QGP does not depend only on the momenta of the 
two subjets, but also on their angular separation $\Delta R$.
To better understand the systematics of that modification, it is instructive to 
study the density of subjets in the ($\log(1/z_g) , \log(1/\Delta R )$) plane, which provides much more information than the individual distributions of $z_g$ or $M_g$. 
Two-dimensional distributions of this type, known as the Lund plane, have
been very useful in the design of new observables for pp collisions at the LHC
(see Ref.~\cite{Dreyer:2018nbf} for a recent discussion) and are becoming a widespread tool for the analysis of jet physics in-medium as well~\cite{Andrews:2018jcm}.
In Section~\ref{sec:lundplane} we shall present Lund plane distributions defined from the two
subjets that first satisfy the Soft Drop condition obtained from our model calculations done with varying $\lres$.

Once the Soft Drop condition is met and two subjets satisfying it have been found, 
the Soft Drop procedure can be iterated, for our purposes with the goal of providing an operational
count of how many hard splittings there are within the original jet.  This can be done in a number of 
ways; we shall employ the iterated Soft Drop (ISD) procedure of Ref.~\cite{Frye:2017yrw}.
In the ISD procedure, the softer of the two subjets found previously is recorded and removed, and the Soft Drop procedure is then applied anew to the harder of the two subjets.  Further declustering and grooming, following the Soft Drop procedure, ensues until the Soft Drop condition (\ref{SoftDropCondition}) is again met, with two new subjets identified.  Again the softer of these two subjets is recorded and removed, and the Soft Drop procedure is applied yet again to the harder one.
These steps are repeated iteratively until the entire Cambridge/Aachen reclustering of the jet has been undone.
From this iterative procedure, the `number of Soft Drop splittings', $n_{\rm SD}$, is defined.
$n_{\rm SD}$ counts the number of times that the Soft Drop condition is satisfied as the hardest branch of the jet is iteratively declustered.  It is a measure (not the only possible choice of such a measure) of the multiplicity of hard splittings within the original jet.
In Section~\ref{sec:nsd} we shall present results for the distribution of $n_{\rm SD}$  from our
model calculations with varying $\lres$.

For an angular ordered shower as in a jet that showers in vacuum, this iterative procedure based upon sequentially undoing the Cambridge/Aachen reclustering of the jet constituents 
starts from the hardest parton early  in the shower that satisfies the Soft Drop condition and
follows its subsequent splittings, in order.
Jets that shower within a droplet of QGP need not be angular ordered, however, for several reasons: (i) because 
the reconstructed jets necessarily include soft particles originating from
the hadronization of the moving plasma -- the wake -- that the jet leaves behind; (ii) because
the jets may include partons from the medium that have been kicked by partons in the jet; and (iii) because partons in the jet that receive a kick may radiate gluons at large angles.  
If the shower is not angular ordered, the sequence of steps in the Soft Drop procedure
will not follow the sequence of steps that occurred during the branching process itself.
This means that the interpretation of Soft Drop observables in heavy ion collisions
is necessarily different than in proton-proton collisions.
It is nevertheless very interesting to measure how such observables are modified
in heavy ion collisions because in any given model for how jets interact with the medium it
is possible to study how sensitive these observables are to various different physical phenomena.
In the following subsections we will explore the effect of the medium on different Soft Drop observables.

\subsection{Number of Soft Drop splittings, $n_{\rm SD}$ }\label{sec:nsd}

\begin{figure}
\centering 
\includegraphics[width=.5\textwidth]{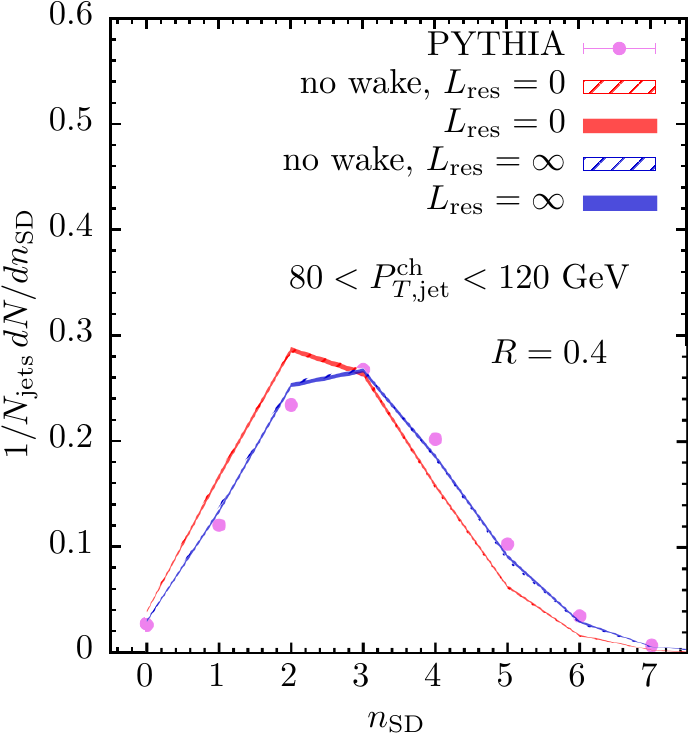}
\caption{\label{Fig:nsd} Distribution of the number of Soft Drop splittings, $n_{\rm SD}$, 
in our hybrid model calculations with 
$\lres=0$ (red bands) and $\lres=\infty$ (blue bands) for 0--10$\%$ central heavy ion collisions
with $\sqrt{s}=2.76$ ATeV. We compare these results to the $n_{\rm SD}$ distribution for jets
in proton-proton collisions that shower in vacuum, obtained from PYTHIA and shown as the purple dots.
The Soft Drop procedure used in the calculation is specified by the grooming parameters in the Soft Drop condition (\ref{SoftDropCondition}), chosen as
 $z_{\rm cut}=0.1$ and $\beta=0$.   The red and blue dashed bands (which are almost identical to the red and blue solid bands) show the results that we obtain upon ignoring particles 
originating from the moving wake deposited by the jet in the plasma; since these particles are soft, they hardly contribute to a groomed observable like $n_{\rm SD}$.  
}
\end{figure}

The first observable whose distribution we shall calculate and present is 
the number of Soft Drop splittings $n_{\rm SD}$. As we have discussed above, this provides one measure of the number of subjets within a jet. The results from our hybrid model calculations are shown in Fig.~\ref{Fig:nsd}. 
This observable
has been recently measured in PbPb collisions at the LHC  by the ALICE collaboration~\cite{Acharya:2019djg}  using what we shall refer to as the flat grooming procedure, with $z_{\rm cut}=0.1$ and $\beta=0$. For this reason, we employ this choice of grooming parameters.
With this choice, soft fragments are groomed in a way that is independent of the angle between the candidate branches.

In Fig.~\ref{Fig:nsd} we present the results of our model calculations for the two extreme values  $\lres=0$ and $\lres=\infty$. These two realizations of the model yield distinct $n_{\rm SD}$ distributions.
The  choice $\lres=\infty$ corresponds to the case where
the QGP medium responds to the jet as if it were just a single energetic colored probe regardless of how the partons within the jet split and shower.   In this case, the $n_{\rm SD}$ distribution is very
similar to what it would have been for jets in vacuum: the jet loses energy overall, but its internal structure is not modified and hence neither is the $n_{\rm SD}$ distribution.
The choice $\lres=0$ corresponds to the case where the QGP medium resolves the 
jet fully, with every parton in the shower losing energy independently as the shower develops and propagates in the QGP medium.
The quenched $n_{\rm SD}$ distribution for these fully resolved jets shifts towards smaller  
$n_{\rm SD}$.
This means that the jets with 
$p_T^{\rm jet}$ between 80 and 120 GeV after quenching tend to have a smaller $n_{\rm SD}$ 
than the jets in this $p_T$-range in the absence of quenching.
This is exactly the behavior that we could have expected for a fully resolved shower.
As long as jet constituents are separately resolved by the medium, 
among jets with the same energy before quenching those which contain more lower energy constituents will
lose more energy than those which contain fewer harder constituents.
That is, as long as jet constituents are fully resolved we should expect that jets with larger $n_{\rm SD}$ lose
more energy than jets with the same energy with smaller $n_{\rm SD}$.
Given that the jet spectrum is a steeply falling function of jet energy, this means that 
the jets that remain in a given $p_T$-range after quenching will be those which lose less energy, meaning those which tend to have smaller $n_{\rm SD}$.

Since the number of splittings increases with the angular width of the jet, we can also summarize this behavior by the statement that wider jets lose more energy, with the consequence that the jets that remain in a given $p_T$-range after quenching tend to be narrower.
This conclusion has been demonstrated previously via analyses of other observables, in 
models of jet quenching that are built upon weakly coupled physics~\cite{Milhano:2015mng}, in holographic models of jet quenching that assume strongly coupled physics~\cite{Rajagopal:2016uip,Brewer:2017fqy}, as well as in our hybrid model~\cite{Casalderrey-Solana:2016jvj}.
What we now see from the present calculation is that this effect goes away 
as the ability of the medium to resolve the splittings within a jet is reduced.
In the extreme limit in which $\lres=\infty$ and 
none of the subjets within a jet can be resolved, jet quenching can only depend on
the energy of the jet, not on its internal substructure, and
jets with the same energy but differing number of splittings and hence differing $n_{\rm SD}$ lose energy similarly. 
 In this case, 
the quenched jet $n_{\rm SD}$ distribution is essentially
identical to its initial distribution, prior to quenching.

Quite unlike what we found for the charged jet mass in Section~\ref{sec:chmass}, the $\lres$-dependence in our calculation of the $n_{\rm SD}$ distribution  is robust in the sense that it shows almost no sensitivity to the soft particles that come from the hadronization of the plasma, including the moving wake created in the plasma by the jet.
We illustrate this via the dashed colored bands in Fig.~\ref{Fig:nsd} which show the $n_{\rm SD}$ distributions that we obtain when we neglect
the contribution from the wake in the plasma, which are almost identical to the results of the full computation shown in the solid colored bands.

The observable $n_{\rm SD}$ serves to make the points that we wish to make perfectly in all respects except one:
the differences between the $n_{\rm SD}$ distributions obtained upon making the two extreme assumptions for $\lres$, namely $\lres=0$ or $\lres=\infty$, are small in magnitude.
This means that although our model study of this observable serves to illuminate the salient physics
very clearly, the sensitivity of this observable to changes in the value of $\lres$ is sufficiently limited that it would be rather a challenge to use experimental measurements of this observable
to constrain the value of $\lres$.
Recent measurements by ALICE \cite{Acharya:2019djg} show a small shift of the quenched $n_{\rm SD}$ distribution toward smaller $n_{\rm SD}$, although the uncertainties in the measurement mean that it is still consistent with the vacuum distribution. Also, as we shall discuss further in Section~\ref{sec:conc}, 
these measurements should anyway not yet be compared quantitatively to our calculations since the measurements have not been unfolded and since we are not able to smear the results of our calculations in a way that incorporates detector effects.
While progress on these fronts together with higher precision measurements of the distribution of $n_{\rm SD}$ could in principle be used to determine the resolving power of the medium,
 in the next Subsections
we will explore other observables which are more sensitive to the dependence of the energy loss on the number of propagating sources of energy loss within a jet, and hence are more sensitive
to the value of $\lres$.

\subsection{\label{sec:zgdist}Momentum sharing fraction, $z_g$}

We turn now to the results we obtain from our model calculations of the $z_g$-distribution, and 
how it is modified for jets in heavy ion collisions that fragment in medium relative to what is seen for jets in proton-proton collisions that fragment in vacuum.
Recall that $z_g$, defined in (\ref{zgdef}), describes the momentum sharing ratio of the two subjets identified by the Soft Drop procedure the first time that the condition (\ref{SoftDropCondition}) is satisfied. 
In the case of high-energy jets from proton-proton collisions, with an appropriate 
choice of normalization the $z_g$ distribution provides direct experimental access to information about the partonic splitting functions that govern how a jet showers in vacuum~\cite{Larkoski:2015lea}; this is a central reason for the interest in measuring this distribution in proton-proton collisions. In medium, however,
because the shower development is not angular ordered and because particles appearing at large angles relative to the jet axis can have multiple origins, there is no reason to expect that the $z_g$ distribution can be
interpreted in this fashion.
We shall see that it is nevertheless very interesting to measure the ratio of  $z_g$ distributions in
heavy ion collisions to those in proton-proton collisions. We shall see that a suitably differential measurement of how these distributions are modified in heavy ion collisions can give
us insights into the degree to which the QGP medium resolves constituents within a jet shower,
and ultimately could be used to constrain the value of the resolution length of QGP.
The other motivation for this analysis is that $z_g$ distributions in heavy ion collisions
have been measured by the CMS \cite{Sirunyan:2017bsd}, STAR \cite{Kauder:2017cvz} and ALICE \cite{Acharya:2019djg} collaborations.

\begin{figure}
\centering 
\includegraphics[width=.9\textwidth]{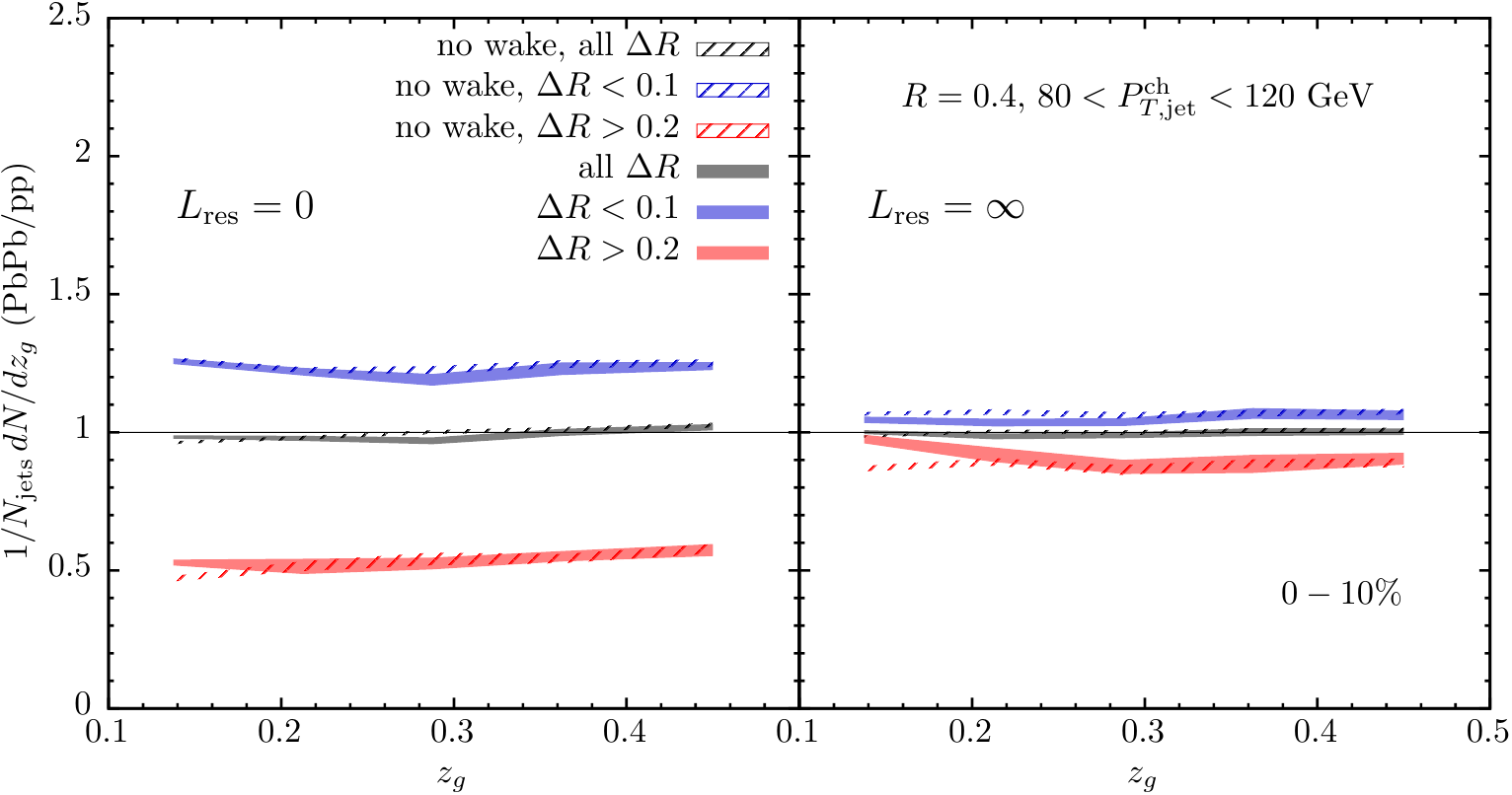}
\caption{\label{Fig:zgdelR} The ratio of the $z_g$ distributions in 0-10$\%$ central PbPb collisions with  $\sqrt{s}=2.76$ ATeV as calculated
in the hybrid model with $\lres=0$ (left panel) and $\lres=\infty$ (right panel) to the  $z_g$ distributions in proton-proton collisions as calculated in PYTHIA.
The $z_g$-distributions for the PbPb and pp collisions are not self-normalized; as in the ALICE analysis of Ref.~\cite{Acharya:2019djg}, each is normalized
to $N_{\rm jets}$,  the number of analyzed jets (reconstructed using the anti-$k_T$ algorithm with radius parameter $R=0.4$; with the transverse momentum in charged particles in the jet in the range 
$80$~GeV$< p_{T,{\rm jet}}^{\rm ch} < 120$~GeV).
In each panel, we show the ratio of $z_g$ distributions for all pairs of subjets found by the
Soft Drop procedure in dark grey, and the ratio of the $z_g$ distributions only for those pairs
of subjets separated in angle by $\Delta R <0.1$ ($\Delta R>0.2$) in blue (red).
In the dashed bands (which are very similar to the corresponding solid bands)
we present the results obtained by ignoring the particles that originate from the moving
wake in the plasma that are reconstructed as part
of the jet. As in Fig.~\ref{Fig:nsd}, since these particles are soft they hardly contribute
to this groomed observable.
}
\end{figure}

The results of our model calculations of the $z_g$ distribution are illustrated in  Fig.~\ref{Fig:zgdelR}. As in our calculations of $n_{\rm SD}$ described in the previous subsection, we  follow ALICE and choose $\beta=0$ and $z_{\rm cut}=0.1$. 
We choose 
not to self-normalize each $z_g$-distribution,
choosing instead to
normalize each distribution by the total number of jets that enter the $p_T$ cuts within the acceptance, which we denote by $N_{\rm jets}$. 
In this way, more information is retained and, in particular, the different relative contributions to $z_g$ of events in which the two subjets have different angular separations can be studied. 
Instead  of presenting the individual $z_g$-distributions, in Fig.~\ref{Fig:zgdelR} we show the ratio of 
the $z_g$ distribution for PbPb collisions as calculated in the hybrid model to that for pp collisions from PYTHIA.
As in our analysis of $n_{\rm SD}$, we present the results of calculations done upon making 
the two extreme assumptions for the QGP resolution length:
$\lres=0$ in the left panel and 
$\lres=\infty$ in the right panel. 
In both panels, the grey band represents the ratio of $z_g$ distributions without any angular cut in the subjet separation, while the blue and red bands represents the ratio of $z_g$ distributions when the two subjets are required to be separated by an angular distance $\Delta R<0.1$ and $\Delta R>0.2$, respectively. 
We see that if we had only looked at the grey bands we would have seen almost no
dependence on $\lres$: if we integrate over subjet pairs with any angular separation, this observable
is not sensitive to whether the medium is or is not able to resolve constituents within jets.
However, 
the comparison between the left and right panels of    Fig.~\ref{Fig:zgdelR} clearly shows that
when we do an analysis that is even somewhat differential in $\Delta R$, focusing on either subjets that are very close together ($\Delta R<0.1$) or those that are more separated ($\Delta R>0.2$) then we find that the ratio of $z_g$ distributions is an observable that is quite sensitive to
the degree to which the medium can resolve jet constituents.

Before continuing on to the next observable that we shall consider, a number of features of the results shown in Fig.~\ref{Fig:zgdelR} warrant discussion.
One characteristic 
feature of all the $z_g$ distribution ratios in the figure (regardless of the angular cut and the value of $\lres$) 
is that the modification of the $z_g$ distribution in heavy ion collisions relative to that in pp collisions is almost completely independent of $z_g$. 
This feature is a natural consequence of our modelling assumption that neither the virtualities nor the $z$ fractions of the splittings in the shower are to be corrected due to quenching effects. 

We note that in the case where $\lres=0$ and the medium is able to resolve all the partons
within the jet shower, careful inspection of the grey curve in the left panel of Fig.~\ref{Fig:zgdelR}
shows that the $z_g$ distribution for the case in which subjet pairs with all values of $\Delta R$  is  very slightly $z_g$-dependent, slightly below one at small $z_g$ and slightly above one at large $z_g$.  This small effect originates from the independent energy loss of the two subjets
that occurs in a medium with $\lres=0$; it is not seen in the case of a medium with $\lres=\infty$.
At the partonic level, this effect is somewhat larger, although not large; hadronization 
reduces it.

Next, we turn to the much more striking feature seen in our calculations with $\lres=0$, in which the medium is able to resolve every parton in the jet shower.
In this case, we see that the ratio of the $z_g$ distribution in heavy ion collisions
to that in pp collisions depends very significantly on how we cut on the angular separation $\Delta R$ between
the subjets identified via the Soft Drop procedure.
If we average over all values of the angular separation $\Delta R$ between the pair of subjets found within an $R=0.4$ jet
via the Soft Drop procedure, the $z_g$ distribution for the jets in medium is almost unchanged
relative to that for jets in vacuum.
However, the blue (red) bands in the left panel of Fig.~\ref{Fig:zgdelR} show that 
if we focus only on narrow (wider) subjet structures, we obtain quite different results. 
The enhanced value of $z_g$ for in-medium subjet pairs that are narrow (small angular separation) tells us that these subjet structures are more common in the ensemble of jets after quenching than in the ensemble of vacuum jets.  Similarly, the suppression of the $z_g$ distribution ratio
for wider subjet pairs shown by the red band tells us that more widely separate subjets
have become less common in the ensemble of jets after quenching.
We conclude that although the shape of $z_g$ distribution itself is hardly modified
by quenching, the shape of the distribution of $\Delta R$, the angular separation between
subjets, must be severely modified by quenching -- as long as the medium can resolve the internal structure of the jets.
This result is again a direct consequence of the observation that wider, larger multiplicity, jets lose more energy in medium -- as long as their internal structure is resolved. 
This  is the same mechanism that leads to a bias towards (narrower) less active jets with
smaller $n_{\rm SD}$ that we saw in Fig.~\ref{Fig:nsd}.

The jet activity, quantified in terms of the number of splittings $n_{\rm SD}$, 
or the narrowness/wideness of the jet, quantified in terms of the angular separation $\Delta R$
between subjets identified via the Soft Drop procedure,
will each be determined by the virtuality of the jet (the jet mass) which can be assigned 
when the first splitting after the hard scattering occurs.  
If this jet mass is large it means that the resulting jet will likely contain subjet structures
separated by a large $\Delta R$. And, to the extent that high virtuality splittings result in 
high multiplicity jets, finding a subjet structure with larger $\Delta R$ should be correlated
with finding a jet with a larger value of $n_{\rm SD}$. (For an explicit check, see Fig.~\ref{Fig:rgvsnsd} in Appendix~\ref{sec:correlation}.)
We can then easily understand why, in the totally resolved limit with $\lres=0$, there is a suppression of the $\Delta R>0.2$ subjet pairs in medium with respect to in vacuum, since such wider jets tend to be more suppressed due to their higher number of effective energy loss sources. Complementarily, the relative contribution of narrow configurations with $\Delta R<0.1$ is enhanced. 

Note that the jet mass defined when the first splitting after the hard scattering occurs 
need not be the same
as the (charged) jet mass as measured in experiment and discussed in Section~\ref{sec:chmass} 
because of all the soft particles from the wake in the plasma that end up reconstructed as
a part of the jet in the final state.  The Soft Drop procedure allows us to get our hands on $\Delta R$, which is 
correlated with the mass of the jet as it was after the first splitting in the jet shower.
In Section \ref{sec:gmass} we shall look at the corresponding groomed jet mass observable itself.

It might also be worth stressing that the shape of the QCD splitting function does not depend on the angle of the emission, given enough phase space, which is largely the case for the first Soft Drop splitting. (See  Fig.~\ref{Fig:zgvsnsd} in Appendix~\ref{sec:correlation} to see this is true for the majority of jets.) This is why we can see a suppression of the wider configurations without any visible modification of the shape of the $z_g$ distribution.

 By comparing the results in the left panel of Fig.~\ref{Fig:zgdelR} that we have discussed above
 with the results in the right panel, which were obtained from our model with $\lres=\infty$ in which
 the medium cannot resolve any substructure within a jet whatsoever we see
 that the strong differences between the large-$\Delta R$ and small-$\Delta R$ $z_g$ distributions seen  in the left panel are characteristic of a medium whose resolution length is small enough that it can resolve structures within jets.
If the medium cannot resolve the constituents of the parton shower as it develops within the medium, then quenching dynamics are insensitive to the jet width, insensitive to the angular separation between subjets $\Delta R$, and insensitive to $n_{\rm SD}$ as we saw in Section~\ref{sec:nsd}. 
The small differences between pp and PbPb $z_g$ distributions that do arise in the right
panel of Fig.~\ref{Fig:zgdelR} 
reflect other effects, such as the change in relative ratios of quark to gluon jets in the quenched jet sample and the reduction of phase space due to quenching. 
  
  The comparison between results obtained from model calculations with the two extreme values of $\lres$ shown in Fig.~\ref{Fig:zgdelR} tell us that we may use measurements of 
  these observables to constrain
  the resolution length of quark-gluon plasma, $\lres$.  We shall defer further discussion of how to do this to Section~\ref{sec:constrainingLres}.
It is pleasing to see, though, that (quite unlike what we found for the charged jet mass in Section~\ref{sec:chmass}) the $z_g$ distribution ratios are all quite insensitive to 
the soft particles coming from the back-reaction of the jet on the medium, namely
from the moving wake in the plasma created by the jet.  This is illustrated
in Fig.~\ref{Fig:zgdelR} by the similarity between the dashed bands (in which the $z_g$ distribution for jets in the medium was computed without any contribution to the jets coming from the wake) and the solid bands with the same color.
We see that for this groomed observable, as for $n_{\rm SD}$, the Soft Drop grooming procedure
has had the desired effect.

We conclude this Section by noting that  since all the different $z_g$ distribution ratios show very little dependence on  $z_g$, meaning that all the $z_g$ distributions have similar shape,
if we had chosen to self-normalize each of the $z_g$ distributions we would not
have seen any of the interesting in-medium effects, and would not have realized that there
are substantial consequences for these observables of the value of $\lres$.
Only if the absolute normalization is kept will it be possible to use 
this observable to discern whether the jet shower is resolved by the medium or whether  the jet behaves in medium as if it were just a single energetic colored probe, losing energy.
The reader may have noticed that, strictly speaking, the norm of these distributions is not under good perturbative control. As shown in 
 Ref.~\cite{Larkoski:2014wba}, the momentum sharing distribution for $\beta=0$ is not an infrared and collinear (IRC) safe quantity. However, while still IRC unsafe, the self 
 normalized $z_g$-distribution {\it is} Sudakov safe~\cite{Larkoski:2015lea} and, as a consequence, is amenable to perturbative computation. 
  We have also tested that there is no problem in practice by repeating the entire present analysis
 upon choosing 
 a small negative value $\beta=-0.05$ instead of $\beta=0$;  this makes the observable IRC safe, and we find that our results change by less than 5\%. However,  to make contact with 
 the already existing measurements of this quantity by the CMS and ALICE collaborations, we chose to present our $\beta=0$ results here. 
That said, a direct comparison between our results for this observable and measurements
at the LHC as of today is not straightforward, and may even be misleading, since  the 
measurements have not yet been unfolded. We will come back to this point in 
Section~\ref{sec:conc}.

\subsection{\label{sec:lundplane}Lund plane}

As we have discussed,
the inclusion of an angular restriction to the momentum sharing fraction ($z_g$) distribution leads to a much larger sensitivity to in-medium physics than the angular-averaged counterparts. As we have already argued,  this observation implies that the modification of the internal structure of in-medium jets does not only depend on the momentum of the fragments but also on their angular distribution. To better understand the systematics of that suppression, it is instructive to 
study the density of subjet pairs, which is to say first hard splittings in the shower, in the ($\log(1/z_g) , \log(1/\Delta R)$) plane and how this
is modified in PbPb collisions, as this provides much more information than the various $z_g$ distributions presented in the previous section. 
This two-dimensional Lund plane distribution has been widely used
in the design of 
new observables for pp collisions~\cite{Dreyer:2018nbf} and is becoming widespread
as a  tool with which to analyze in-medium jet physics also~\cite{Andrews:2018jcm}.

\begin{figure}
\centering 
\includegraphics[width=0.99\textwidth]{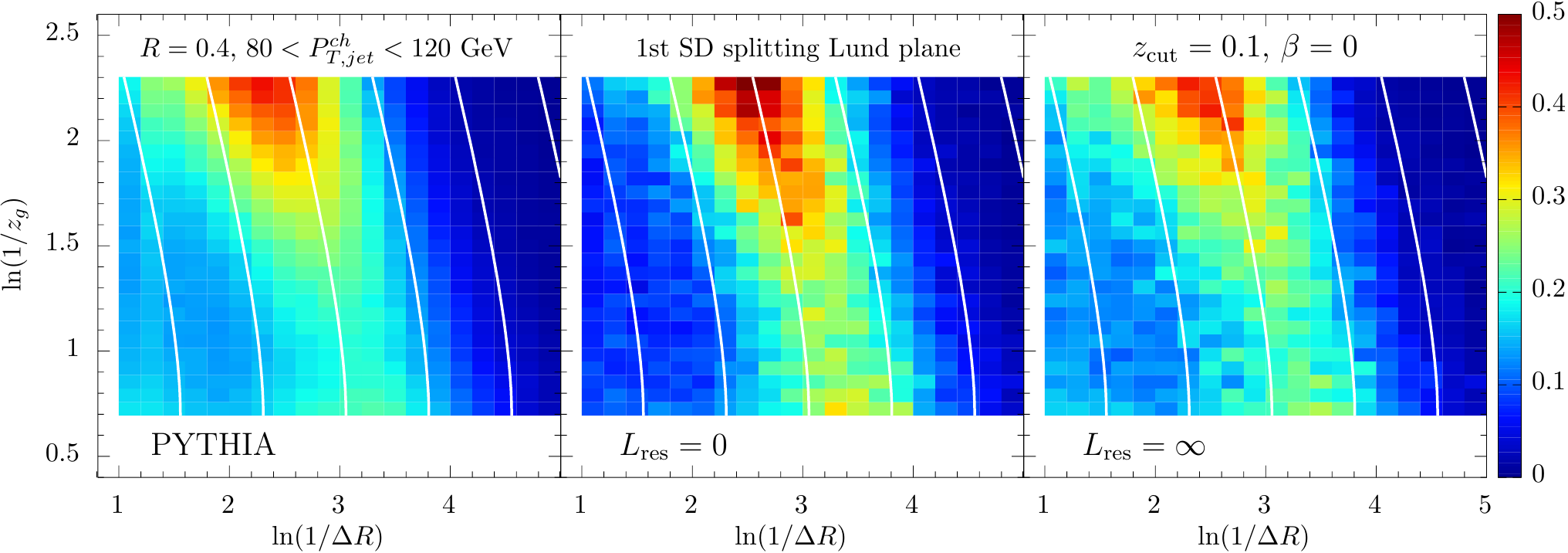}
\caption{\label{Fig:glundInd} The Lund plane defined by the first pair of subjets
found via the Soft Drop procedure that pass the Soft Drop condition (\ref{SoftDropCondition}).
The three panels show this Lund plane for jets in vacuum as described by PYTHIA, and
for jets in a medium with $\lres=0$ (middle panel) and $\lres=\infty$ (right panel) as described by the
hybrid model at $\sqrt{s}=2.76$ ATeV.
Color indicates the density in the 
($\log(1/z_g) , \log(1/\Delta R)$) plane corresponding to the probability of finding
such subjets with a given $z_g$ and $\Delta R$.
We have used the flat grooming procedure ($\beta=0$ and $z_{\rm cut}=0.1$) 
and the distributions are normalized to the total number of analyzed jets, $N_{\rm jets}$, within the cuts 
$80<P^{\rm ch}_{T, \rm jet}<120$ GeV and $|\eta|<0.9$. The jets were
 reconstructed with anti-$k_t$ radius $R=0.4$.
White curves correspond to contours of constant log(1/($M_g/p_{T,g}$)), where $M_g$ is the groomed mass and $p_{T,g}$ is the groomed transverse momentum of the jet.}
\end{figure}

In Fig.~\ref{Fig:glundInd} we show our model calculations for the Lund
plane distribution defined by the  first pair of subjets found during the Soft Drop declustering procedure that pass the Soft Drop condition (\ref{SoftDropCondition}).  
This Lund plane distribution is constructed from  the angular separation $\Delta R$ and the 
momentum fraction $z_g$ of this pair of subjets.
This distribution depends on the 
Soft Drop parameters; the results presented in Fig.~\ref{Fig:glundInd} are for the flat grooming procedure ($\beta=0,\, z_{\rm cut}=0.1$). 
To make contact with ALICE measurements, the total transverse momentum of charged particles in the jets that populate these distributions 
are restricted to $ 80$~GeV $< p^{\rm ch}_{T\, \rm jet} < 120$~GeV and their pseudorapidity is within  $\left| \eta\right| <0.9$. In this figure, the left panel shows the distribution for our reference vacuum computation, given by PYTHIA, and the middle and right panels correspond to the two extreme assumptions for the in-medium resolution parameter, $\lres=0$ and $\lres=\infty$ respectively. 
Each Lund plane distribution is normalized to the number of jets that pass the cuts, $N_{\rm jets}$.
The $z_g$ distributions presented in Fig.~\ref{Fig:zgdelR}
may be obtained by integrating these two-dimensional distributions in the corresponding intervals of $\Delta R$.

The comparison of the different distributions shown in  Fig.~\ref{Fig:glundInd} 
allows us to understand important features of in-medium jet evolution. 
Let us start our discussion by comparing the totally unresolved case, $\lres=\infty$, with the reference PYTHIA computation. 
As we saw for the momentum sharing fraction in Section~\ref{sec:zgdist}, the density of fragments in the entire Lund plane for unresolved in-medium jets is almost identical to that for unmodified vacuum jets. Extending the discussion of the previous section, these results show that in the totally unresolved case (in which the medium is unable to resolve any structure within a jet)
the distribution of the primary Soft Drop subjets are almost the same after quenching
as they were in vacuum.
This is certainly expected,
since in this case in-medium energy loss does not 
alter the relative weights of the different splittings since it cannot `see' them.

In contrast, when the jet is totally resolved, meaning that $\lres=0$ and the medium is such that each parton in the developing jet shower loses energy independently,
the distribution of the primary Soft Drop subjets in the Lund plane is clearly modified.
As shown by comparing the middle and left panels of   Fig.~\ref{Fig:glundInd}, for a fixed value of the momentum sharing fraction $z_g$ the density of splittings is shifted towards smaller angles (or larger values of $\log (1/\Delta R)$): the red shifts to the right.
This is a direct manifestation of the observation that when the resolution length of the medium
is short enough that substructures within a jet are resolved, jets with larger $\Delta R$ and more
constituents lose more energy.
Hence, the observed distribution after quenching is biased toward subjets with narrower angular separation $\Delta R$ in a medium with $\lres=0$. 
To better illustrate this 
effect, in all panels of  Fig.~\ref{Fig:glundInd} we have shown contours of fixed $\log(1/(M_g/p_{T, g}$)), where $M_g$ 
is the groomed mass and $p_{T, g}$ is the groomed transverse momentum of the jet, as defined from Eq.~(\ref{Pgdef}).
Using simple kinematics, and neglecting the effect of rapidity, the quantity $M_g/p_{T, g}$ is 
well approximated by the relation 
\be
\frac{M_g^2}{p_{T, g}^2} \simeq  z_g(1-z_g) \Delta R^2\ ,
\ee
which we have used to draw the white contours in Fig.~\ref{Fig:glundInd}.
This relationship provides an illustration, in the present context, of the relationship
between the ratio of jet mass to jet transverse momentum (here groomed, in both cases) 
to the angular width of the jet (here the angular separation between the two subjets).
As is clearly seen in the distributions displayed in Fig.~\ref{Fig:glundInd}, in the case where the medium has the ability to fully resolve the partons in a jet shower 
the effect of quenching is a shift toward smaller values of $M_g/p_{T,g}$, which is to say a shift
toward narrower groomed jets.  This arises because narrower jets contain fewer active fragments, and it is the wider jets with more active fragments that lose more energy. 
(The reader may wonder whether this could instead be described as an effect due to formation time, with jets that form earlier losing more energy --- and also being wider.  We show
in Appendix~\ref{sec:formationtime} that this is not the right interpretation: any effect of
the formation time on the degree of energy loss is much less significant than the effect of
how many active fragments a jet has and how wide it is.)

In addition to the red ridge shifting to the right, 
the second significant difference that we see in the middle panel of Fig.~\ref{Fig:glundInd} (relative to the left-panel) is that the bottom-left corner of the Lund plane has become lower (bluer), and the red ridge has become higher (redder).  The depletion in the bottom-left corner 
illustrates the fact that if the medium can resolve 
the internal structure of jets then
jets with large $M_g$ (and large $\Delta R$) are less common after 
quenching, because they lose more energy.  
When, because of energy loss, there is no sufficiently hard pair of subjets 
with a splitting that would put them in the bottom-left corner for the Soft Drop algorithm to find,
the Soft Drop algorithm continues, and it becomes more likely that he first subjet splitting that
passes the Soft Drop condition (\ref{SoftDropCondition}) is one of the more common
splittings that populate the red ridge.
For this reason, the depletion in the bottom-left corner also has the effect of pushing
the red ridge higher, making it more red as is seen in the middle panel of Fig.~\ref{Fig:glundInd}.

\begin{figure}
\centering 
\includegraphics[width=1\textwidth]{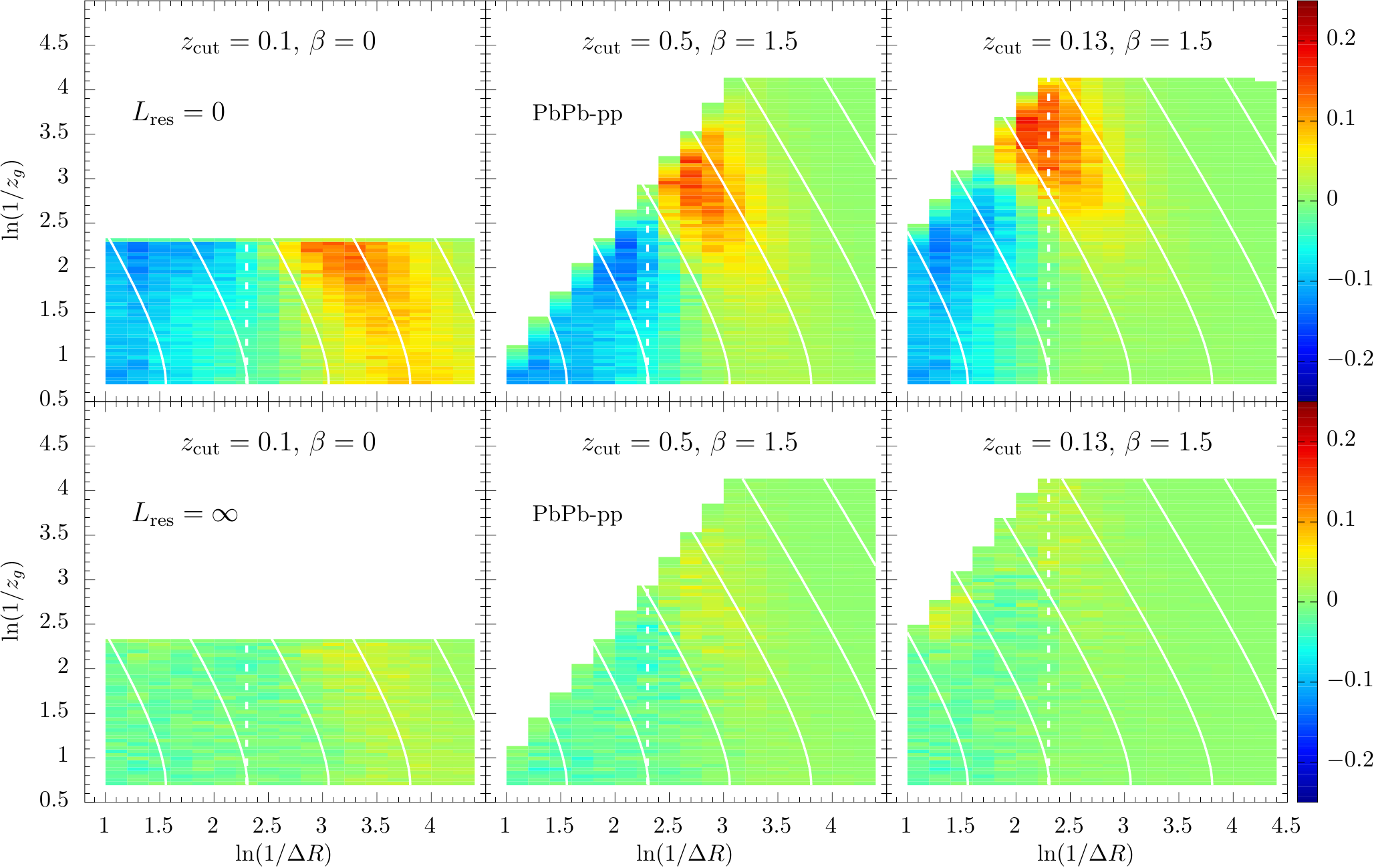}
\caption{\label{Fig:glund} Differences between medium and vacuum for the first Soft Drop splitting Lund planes at $\sqrt{s}=5.02$ ATeV.  Upper row corresponds to $\lres=0$ and lower row to $\lres=\infty$. First column used the flat grooming parameters, the middle column the core grooming parameters, and the right column the soft-core parameters; see text for details. 
In each case, color indicates the difference between the density in the Lund plane for in-medium jets and the density in the Lund plane for vacuum jets.
The individual Lund planes have been normalized to the total number of jets within the cuts, which are $140<p_{T}^{\rm jet}<300$ GeV and $|\eta|<1.3$. The jets were reconstructed with anti-$k_t$ radius $R=0.4$. 
Note that the top-left panel here is, in essence, the difference between the middle and left panels of Fig.~\ref{Fig:glundInd} (not exactly, since the cuts are different)
and the bottom-left panel here is, again in esssence, the difference between the right and left panels of Fig.~\ref{Fig:glundInd}
White solid curves correspond to contours of constant $\log(1/(M_g/p_{T, g})$), where $M_g$ is the groomed mass and $p_{T, g}$ the groomed transverse momentum of the jet. The vertical dashed white line represents an angular separation of $\Delta R = 0.1$.}
\end{figure}

To better 
highlight the differences in the splitting patterns between vacuum jets and in-medium jets, in the two left panels of Fig.~\ref{Fig:glund} we show the 
difference between densities in the Lund plane of Fig.~\ref{Fig:glundInd} for in-medium jets 
and that for vacuum jets, both for the medium with $\lres=0$ and for the medium with $\lres=\infty$.
In these left panels, we use the flat grooming parameters $z_{\rm cut}=0.1$ and $\beta=0$
in the Soft Drop condition (\ref{SoftDropCondition}).
For reasons that will become apparent shortly, 
in the middle and right panels we repeat this calculation 
for two other choices of how we do the Soft Drop grooming, namely choosing grooming parameters $z_{\rm cut} = 0.5$ and $\beta=1.5$, which we call ``core'', in the middle panels and $z_{\rm cut}=0.13$ and $\beta=1.5$, which we call ``soft-core'', in the right panels. Because of these 
different choices of grooming parameters, what gets dropped before the Soft Drop procedure finds a pair of subjets that pass the Soft Drop condition (\ref{SoftDropCondition}) will differ, and hence so will the kinematics of the pairs of subjets identified by the Soft Drop procedure and used to populate the
Lund plane distributions.
Both configurations with non-vanishing $\beta$ allow softer subjets than in the flat grooming procedure, provided that they are separated by a sufficiently small $\Delta R$, while they will tend to reject soft, large angle, structures.

 We see from the lower three panels in Fig.~\ref{Fig:glund} that there is no significant difference
 in the Lund plane density for jets in medium relative to those in vacuum if the medium has $\lres=\infty$, meaning that it cannot resolve any structure within a jet.  This observation applies
 equally to all three choices of grooming parameters. 
  This is, once again, a manifestation of the fact that if the medium has $\lres=\infty$ then 
jet energy loss cannot depend on any jet structure, and no groomed jet observable 
can show significant differences between vacuum and in-medium jets. 

Turning now to the case of a medium with $\lres=0$ that can resolve all the partons in a jet shower,
for all three choices of the grooming parameters the upper panels in Fig.~\ref{Fig:glund}
show clear differences in the Lund plane densities of subjet pairs for jets in medium relative to jets in
vacuum.
Let's start our discussion with the flat grooming procedure, displayed in the top-left panel. 
This panel highlights the features that we already discussed when we compared
the middle and left panels of Fig.~\ref{Fig:glundInd}
and at the same time explains the 
systematic dependence observed in the $z_g$ distribution at different values of $\Delta R$ shown Fig.~\ref{Fig:zgdelR}.
In the top-left panel of  Fig.~\ref{Fig:glund}, we see a clear suppression of subjet pairs for large angular separations, below $\log (1/\Delta R)\sim 2.3$,
 and a clear enhancement above this angle.
In Fig.~\ref{Fig:zgdelR} when we 
select subjet configurations with $\Delta R<0.1$ we are selecting
the region of the Lund plane to the right of $\log(1/\Delta R) \sim 2.3$, which would correspond to 
cutting in such a way that we include the regions of phase space where the Lund plane density
is most enhanced.
In contrast, when we select subjet configurations with $\Delta R >0.2$, which corresponds to
integrating over regions of the Lund plane below $\log(1/\Delta R) \sim 1.6$, 
we are instead capturing most of the phase space where the Lund plane density is most
depleted. This is why the selected choices represented in Fig.~\ref{Fig:zgdelR}, which were also the ones chosen by ALICE \cite{Acharya:2019djg}, are  close to the optimal for the purpose of discriminating  between $\lres=0$ and $\lres=\infty$, at least for the flat grooming procedure.

Another feature that can clearly be inferred from the upper panels of Fig.~\ref{Fig:glund} is the effect of a limited angular resolution on groomed subjet jet measurements. 
The calorimetric measurements performed by CMS~\cite{Sirunyan:2017bsd,Sirunyan:2018gct} restrict the minimum opening angle between the subjets to be 
$\Delta R>0.1$. This restriction means keeping only the region of the Lund planes to the
left of the dashed vertical white lines 
in  Fig.~\ref{Fig:glund}. We see from the Figure that when either the flat (top left) or the core (top middle) grooming procedure is employed, any groomed in-medium observable 
computed with this restriction will be significantly suppressed relative to that for jets in vacuum in the case where the medium has $\lres=0$.
In addition, since the Lund plane density distribution does not show much $z_g$-dependence in this region, the suppression will be at most weakly $z_g$-dependent. 
And indeed, for the flat grooming procedure this is what we found in Fig.~\ref{Fig:zgdelR}.
In contrast, by inspecting the region of the top-right panel of Fig.~\ref{Fig:glund} that lies 
to the left of the dashed vertical white line we see that if we choose the soft-core grooming
procedure 
the density of splittings within this region if the Lund plane shows more structure. 
We therefore expect 
that the medium-dependent dynamics of
some Soft Drop groomed observables will exhibit sensitivity to $z_g$ if we choose soft-core grooming.
We will see an explicit example in Section~\ref{sec:gmass}.

Observation of the subtracted Lund plane 
density distributions in Fig.~\ref{Fig:glund} suggests that if we want to maximize the sensitivity to in-medium effects, a better way to slice this Lund plane may be
to consider fixed values of the ratio $M_g/p_{T, g}$. As in Fig.~\ref{Fig:glundInd},
fixed values of $\log(1/(M_g/p_{T, g})$) are represented by the white lines displayed in those planes. We see that the regions where the density of in-medium 
splittings is either enhanced or suppressed largely lie within intervals of this ratio. 
This suggests that the groomed mass distribution should
exhibit clear differences between  in-medium showers and vacuum showers if the medium has $\lres=0$. 
As we have just described, though, such
measurements made with the restriction that $\Delta R>0.1$ will not exhibit much structure if we choose either the flat or the core grooming procedures.
On the contrary, if we choose the soft-core grooming procedure we expect
a significant enhancement of  the density of splittings
with some (small) values of the groomed jet mass $M_g$
and a significant suppression for other (large) values of $M_g$.
We shall see these expectations realized in the next Section.

\subsection{Groomed jet mass}
\label{sec:gmass}

\begin{figure}
\centering 
\includegraphics[width=1\textwidth]{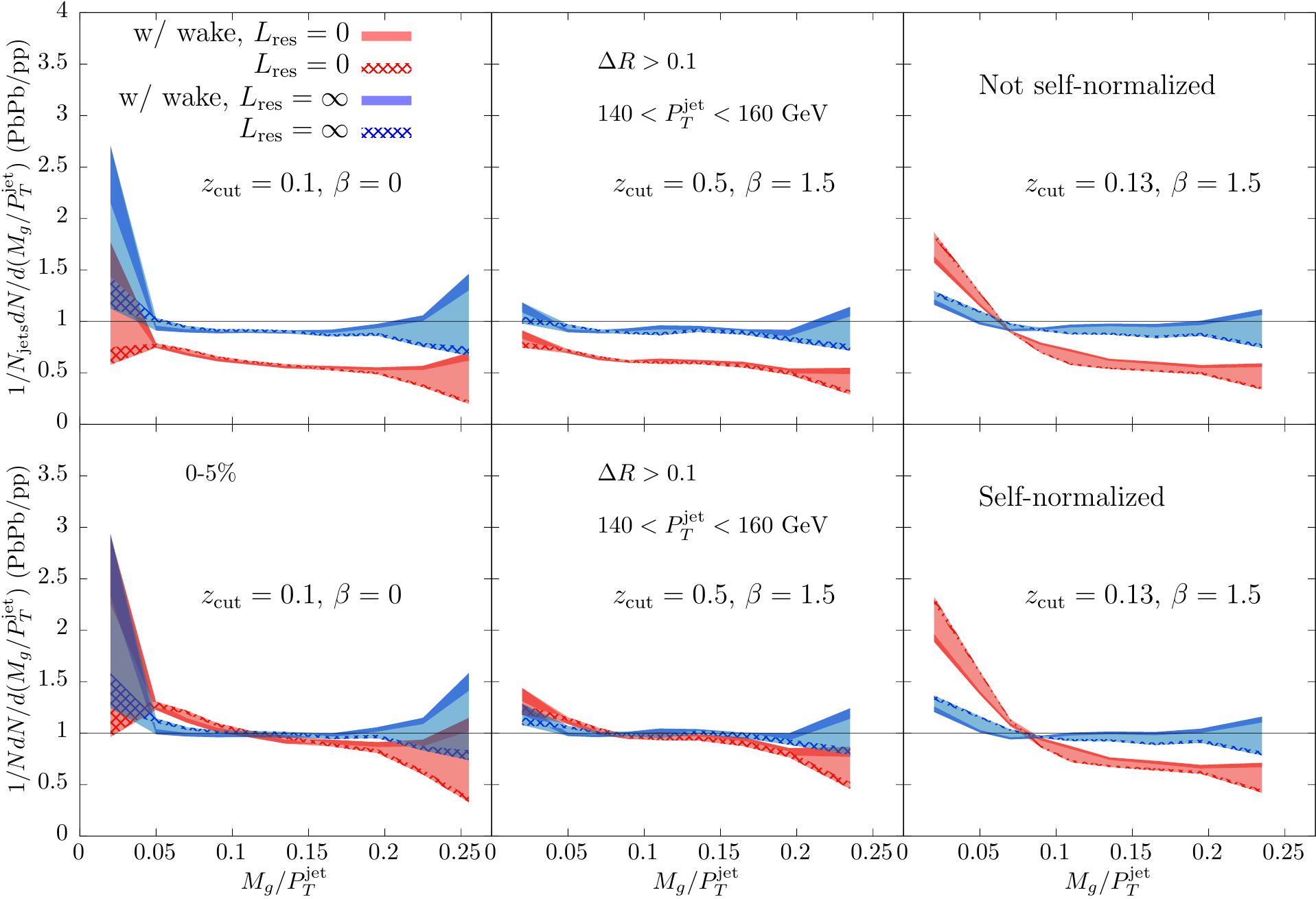}
\caption{\label{Fig:gmass} 
Ratios of the distribution of $M_g/p_T^{\rm jet}$ in PbPb collisions over that in pp collisions at $\sqrt{s}=5.02$ ATeV.
(We use $p_T^{\rm jet}$ rather than the groomed $p_{T,g}$ in the  denominator only because
that is what the experimentalists have chosen to present in Ref.~\cite{Sirunyan:2018gct}.) 
In the three top panels, we normalize each PbPb and each pp distribution to the total number of jets found within the cuts
in that analysis; in the three bottom panels, we self-normalize each PbPb  and each pp distribution.
Red (blue) curves correspond to model calculations with a medium that has $\lres=0$ and can fully resolve the parton shower
($\lres=\infty$ and sees the jet as a single unresolved object).
In solid curves we show the full results of our model calculation which include particles
coming from the wake in the medium;  in dashed curves we show results which don't include those soft particles. To facilitate the visualization of the contributions of the wake to this observable, 
we have colored the space between the solid and dashed curves with the same choice of $\lres$.
Hence, the width of each colored band shows the effect of the wake on this observable and the
difference between red and blue shows the effect of varying $\lres$.
}
\end{figure}

Motivated by the results of our Lund plane measurements described in the previous Section, here we present our model calculation for distributions of the $M_g/p_T^{\rm jet}$ ratio, 
that is the number of 
jets that pass the Soft Drop condition in a given interval of that ratio. In Fig.~\ref{Fig:gmass} we show the ratio of the distribution that we calculate in the hybrid model 
for jets in PbPb collisions  that create a medium
with $\lres=0$ (red) or $\lres=\infty$ (blue) to the distribution that we calculate in PYTHIA for jets in pp collisions. 
In the upper row of the figure we choose to normalize the distributions in each calculation by the total 
number of jets selected for the analysis, $N_{\rm jets}$.
 In the lower row, we present the same data but we choose to self-normalize the individual distributions for each system before we take the ratio of PbPb to pp. In other words, we normalize each distribution by the number of jets that pass the Soft Drop condition and the angular cut which, following the CMS analysis in Ref.~\cite{Sirunyan:2018gct}, we have chosen to be $\Delta R>0.1$.
In both rows, the first panel corresponds  to the flat grooming procedure, the middle to the core grooming procedure, and the right panel to the soft-core grooming procedure.
 Results for the two extreme choices of $\lres$ are presented in each plot, 
  with solid lines representing the full calculation (including medium response) and with dashed lines the result of not including the particles coming from the wake in the medium. 
    To visualize the small sensitivity of this observable to the particles coming from the medium, the difference between the solid and dashed curves is shaded, forming colored bands.

 As we by now expect for a groomed observable, this groomed jet mass distribution for the most part
 exhibits little sensitivity to the soft particles coming from the medium whose momenta are correlated with the jet
 because of the wake that the jet creates in the medium.
 This is in stark contrast to the charged jet mass distribution that we analyzed in Section~\ref{sec:chmass}, where we showed that 
 these soft dynamics significantly alter the charged jet mass distribution. 
 
 While the groomed jet mass distribution is largely insensitive to medium response, we note that
 the contribution of  this physics to this observable does depend on the choice of Soft Drop grooming parameters.  When we choose the soft-core or core grooming procedures,
 the observable is affected by the medium response less than when we choose the flat grooming
 procedure
 since, as shown in Fig.~\ref{Fig:glund}, 
 when $\beta=1.5$ the top-left region of the Lund plane is excluded because it does not 
 satisfy the Soft Drop conditon (\ref{SoftDropCondition}) and this is the region
 where the particles coming from the wake in the plasma make the most significant contribution. For the case with the flat grooming procedure,
the sensitivity 
 to the particles coming from the wake in the medium is greatest at the 
 smallest and largest values of the groomed jet mass. These bins are barely populated,
 meaning that the big enhancement in the sensitivity to the wake that we see there comes
 from taking the ratio between two small numbers.
 The observed enhancement of 
the large jet mass end of the distribution is a natural expectation for the consequence
of including additional soft particles at large angles coming from the wake in the medium. 
 The enhancement observed at the small jet mass end of the distribution 
 is also an effect of the wake in the medium, although in this case the reason
 is less obvious --- and is an artifact. As in the case of our calculation of
 the charged jet mass as discussed in Section~\ref{sec:chmass}, in our treatment
 of the particles coming from this wake we must introduce ``negative particles'' which
 mimic the effect of over-subtraction of a homogeneous background~\cite{Casalderrey-Solana:2016jvj}. One artifact of this is that a small fraction of jets have a negative mass squared,
 as we noted in Section~\ref{sec:chmass}. A second is that the groomed jet mass distribution
 is shifted very slightly to the left.  This artifact is inconsequential everywhere in the distribution except at very small values of the groomed  jet mass.  There, the magnitude of the distribution in pp collisions is very small which means that when the distribution in PbPb collisions is pushed even slightly to the left, the PbPb/pp ratio of distributions rises artificially.

Quite unlike in the case of the charged jet mass distribution of Section~\ref{sec:chmass},
because the groomed jet mass observables are relatively insensitive to the soft particles 
coming from the wake in the medium whose momenta are correlated with that of the jet
we are able to use these observables to discriminate between 
the two extreme assumptions for the value of the resolution length of QGP that we have investigated to this point.   (We shall look at other values of $\lres$ in Section~\ref{sec:constrainingLres}.) 
As is clear from the upper row of Fig.~\ref{Fig:gmass}, when we normalize the groomed jet mass distributions 
by  $N_{\rm jets}$ they are well able to discriminate between $\lres=0$ and $\lres=\infty$ for
any of the three choices of grooming parameters that we have considered.
From the lower row of Fig.~\ref{Fig:gmass}, we see that if we self-normalize each groomed jet mass
distribution then in the case of the flat and core grooming procedures we lose this discriminating power, which means that in these two cases the discrimination that is manifest in the upper row of the figure comes almost entirely from the normalization, which is to say from the number of jets that
pass the Soft Drop condition and the angular cut.
This can easily be understood by noting that in the region $\Delta R>0.1$, to the left of the
dashed white lines in Fig.~\ref{Fig:glund}, the Lund plane distributions for the case of 
a medium with $\lres=0$ presented in that figure
show a depletion of splittings relative to what would be seen in vacuum and, by extension, in a medium with  $\lres=\infty$.  And, other than this depletion there is relatively little structure 
seen in the Lund plane density in this region.
In contrast, our newly proposed soft-core grooming procedure
exhibits a non-trivial $M_g/p_T^{\rm jet}$ dependence in the $\Delta R>0.1$ region of the top-right Lund plane in Fig.~\ref{Fig:glund},
which  translates directly into the conclusion that we reach from the two right panels of
Fig~\ref{Fig:gmass}: the  groomed jet mass distribution (distribution of $M_g/p_T^{\rm jet}$) 
can be used to discriminate between $\lres=0$ and $\lres=\infty$ whether the distributions are 
self-normalized or normalized by $N_{\rm jets}$.

As we noted in Section~\ref{sec:zgdist} for the absolute normalization 
of the distribution of the momentum fraction $z_g$,  the absolute normalization of the groomed mass distribution may not be a reliable observable
 since the groomed jet mass distribution with $\beta\geq 0$ is not IRC safe.
 This fact implies that this observable is, in principle, not well defined in perturbation theory, since it  becomes very sensitive on the non-perturbative regulator of the 
 collinear divergence. In our PYTHIA based analysis, this is the virtuality that terminates the jet shower.  
  As we showed in the analysis of the $z_g$ distribution, this is not a problem in practice for $\beta=0$, at least for the range of groomed jet masses in which the contribution coming from the 
  wake in the medium is small. 
  We have explicitly checked that for intermediate values of  $M_g/p_T^{\rm jet}$  where this contribution is small, the distribution with $\beta=0$ differs by less than 4\% from 
  the IRC safe distribution obtained by choosing $\beta=-0.05$.
  On the contrary, at large or small values of $M_g/P_T^{\rm jet}$ at the edges of its distribution,
   this  variation can be as much as 30\%. This is another way to see that these edge regions are sensitive to soft dynamics. 

 For $\beta=1.5$, there is no IRC setup to which we can directly compare our results.
Nevertheless, 
the self-normalized $M_g/p_T^{\rm jet}$ distribution is, once again, Sudakov safe.
This means that this observable is amenable to perturbative analysis.
It is therefore pleasing that this distribution has such discriminating power vis a vis $\lres$.
 In addition, although the ratio of self-normalized distributions 
obtained using the flat and core grooming procedures do not have this discriminating power, they
have the virtue that they connect to 
 existing measurements of the the groomed mass distribution presented by CMS \cite{Sirunyan:2018gct}. 
So, while the self-normalized distributions obtained using these two grooming procedures
may not be well-suited to constraining the resolution length of the medium, since most of their discriminatory power relies on the norm, our newly introduced soft-core grooming procedure provides us with a robust observable with which to learn about the resolution length of QGP. 
In the next Section, we will explore the sensitivity of this observable to particular choices of
the value of $\lres$.

\subsection{Constraining the resolution length of quark-gluon plasma}\label{sec:constrainingLres}

From the results of the previous subsections, we have identified two robust 
observables that are effective in discriminating between our two extreme assumptions for 
the resolution length of the medium, $\lres=0$ and $\lres=\infty$.  The first such
observable is the $z_g$ distribution in the case where we use the flat grooming procedure.
The second such observable is the groomed mass distribution in the case where we use the soft-core grooming procedure.
In this Section, we will explore our model predictions for one particular intermediate and realistic
value of the resolution length, namely $\lres=2/\pi \,T$, 
in order to better assess the power of these observables to constrain the value of $\lres$.

First, though, an important note.  Up to here we have only treated 
the two extreme values of $\lres$, and in both these cases our calculations
are independent of how the effects of a finite resolution (as opposed to perfect resolution or no resolution) are modelled.  In both  cases that we have treated, the model calculation that we do
depends on how energy loss is modelled but there are no additional model assumptions
related to implementing $\lres$ needed.

Now that we wish to investigate a finite value of $\lres$ we must decide how to model its
effects.  We shall use the particular implementation of resolution effects developed and described in detail in Ref.~\cite{Hulcher:2017cpt}.
In addition, as in that paper and as is physically reasonable, we shall assume
that $\lres \propto 1/\mu_D$, where $\mu_D$ is the Debye mass and hence $1/\mu_D$ is
the screening length of the medium.
The screening length of QGP can be thought of, somewhat loosely, as the minimal separation between two static test color charges such that there is
enough QGP between them so that the two charges are independent of each other, meaning that they exert no force on each other.  In our case, the resolution length of QGP is the minimal
separation between two color charges moving through the QGP at  ultrarelativistic speeds such that
there is enough QGP between them so that the two charges are independent of each other,
meaning that they lose energy independently.  These two length scales need not be identical, but
it is reasonable to assume that they are proportional.  This assumption means
that $\lres \propto 1/(g T)$ if the gauge coupling $g$ is weak, and $\lres \propto 1/T$ if the
gauge coupling is strong. By comparing weak and strong coupling expressions for
$\mu_D$, the authors of Ref.~\cite{Hulcher:2017cpt} argue that it is reasonable to
guess that $1/(\pi T) \lesssim \lres \lesssim 2/(\pi T)$ in QGP.  
In this Section we shall explore our model predictions for one representative value of $\lres$, namely $\lres=2/(\pi T)$.
We leave a more refined exploration of its value for future work.

\begin{figure}
\centering %
\includegraphics[width=1\textwidth]{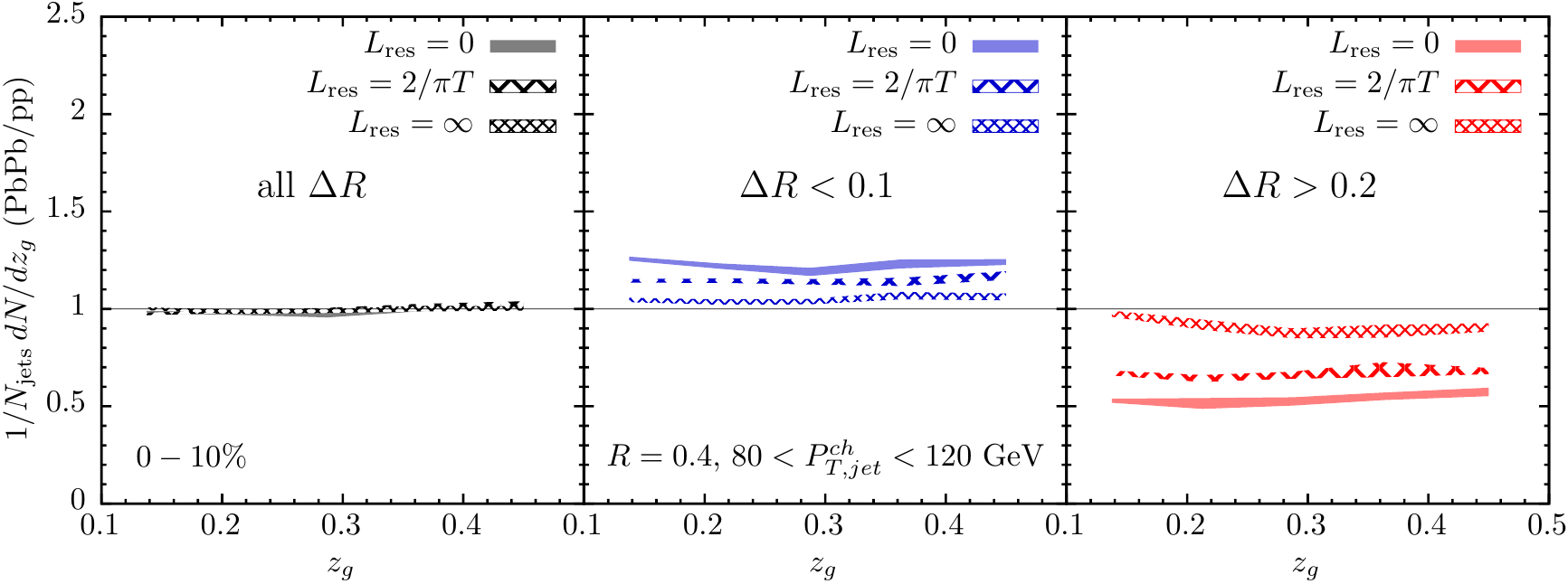}
\caption{\label{Fig:zgLresfinite}  Results from the hybrid model calculation of 
the ratio between the $z_g$ distribution in PbPb and pp collisions, with different choices of $\lres$, depicted via bands with different fillings. Each panel shows a different choice for the 
cut on the angular separation $\Delta R$ between the two groomed subjets.
Each curve has been normalized
to the total number of jets, $N_{\rm jets}$.}
\end{figure}

Results from our model calculation of the $z_g$ distribution for jets
in a medium with $\lres=2/(\pi T)$ 
obtained using the flat grooming procedure 
are presented in Fig.~\ref{Fig:zgLresfinite}. 
We show the ratio of the $z_g$ distribution in PbPb to that in pp collisions as determined by PYTHIA. 
Just as we found for $\lres=0$ and $\lres=\infty$ 
in Section~\ref{sec:zgdist}, 
if we average over all values of the angular separation between the subjets the $z_g$ distribution, 
see the left panel of Fig.~\ref{Fig:zgLresfinite}, is largely independent of 
the value of $\lres$.
On the contrary, when 
we restrict the angular separation between the two Soft Drop subjets either to $\Delta R < 0.1$ or
to $\Delta R >0.2$,
the total number of jets that satisfy the Soft Drop condition and 
pass the cuts in PbPb and in pp collisions greatly differ, 
and depend signficantly on $\lres$, as illustrated by the clear separation between the curves plotted
in the middle and right panels of Fig.~\ref{Fig:zgLresfinite}.
The results of our simulation with 
 $\lres=2/\pi T$  are  clearly distinct from the two extreme values, since the separation between the curves is larger than the theoretical uncertainty of our model computation. 
 We note that the $z_g$ distributions for the medium with $\lres=2/(\pi T)$ are somewhat
 closer to those for the medium with $\lres=0$ that can resolve every parton in the shower than to those for the medium with $\lres=\infty$ that cannot resolve any substructure at all.
This indicates that for values of $\lres$ between $1/(\pi T)$ and $2/(\pi T)$ a significant
fraction of the partons within a jet shower are separately resolved.
The proximity of the $\lres=0$ and $\lres=2/(\pi T)$  results
serves as a gauge of the level of both theoretical and experimental precision needed to use this observable to constrain the resolution length of the plasma, and in particular to establish that it is nonzero. Establishing that some jet substructure is resolved, meaning establishing that $\lres$ is not infinite, should be easier.  In the case of this observable this could be done by establishing that
the $z_g$ distribution ratio obtained using the flat grooming procedure
is significantly different from, and greater than, one if the cut $\Delta R < 0.1$ is imposed
and is significantly different from, and less than, one if the cut $\Delta R>0.2$ is imposed.

 \begin{table}
   \centering
    \begin{tabular}{| l | l | l | l |}
    \hline
     & $\Delta R$\, > 0.0 & $\Delta R$\, < 0.1 & $\Delta R$\, > 0.2 \\ \hline
    PYTHIA & 0.9729(2) & 0.5757(7) & 0.1730(4) \\ \hline
    $\lres=0$ & 0.9599(8) & 0.710(4) & 0.092(2) \\ \hline
    $\lres=2/ \pi T$ & 0.9633(8) & 0.660(3) &  0.115(2) \\ \hline
    $\lres=\infty$ & 0.969(1) & 0.603(3) & 0.161(2)\\
    \hline
    \end{tabular}
    \caption{\label{Fig:table} The number of selected jets 
    as a fraction of the total number of jets $N_{\rm jets}$ within the experimental acceptance.
    The number of selected jets is $N_{\rm jets}$ minus the number of jets that never pass the Soft Drop condition (\ref{SoftDropCondition}) minus the number of jets that pass the Soft Drop condition
 but where the two subjets identified after the Soft Drop grooming procedure are separated by an angle $\Delta R$ that does not meet the criterion specified for each column of the table.  
 The rows of the table give the number of selected jets as a fraction of $N_{\rm jets}$ for
 jets in vacuum as calculated in PYTHIA and for jets in PbPb collisions as calculated in the hybrid model at $\sqrt{s}=2.76$ ATeV with three different values
 of the resolution length of the medium.}
\end{table}

Since all the $z_g$ distributions presented in Fig.~\ref{Fig:zgLresfinite} are approximately independent of $z_g$, a simple way to characterize the difference between these curves is to determine the fraction of the total number of selected jets for each configuration. These are tabulated in Table~\ref{Fig:table}. 

We comment on (the challenges associated with) comparisons to present $z_g$ data in the next Section, where we shall also argue that present data disfavor $\lres=\infty$, which is to say they indicate that the quark-gluon plasma produced in heavy ion collisions does resolve some jet substructure.

\begin{figure}
\centering %
\includegraphics[width=0.8\textwidth]{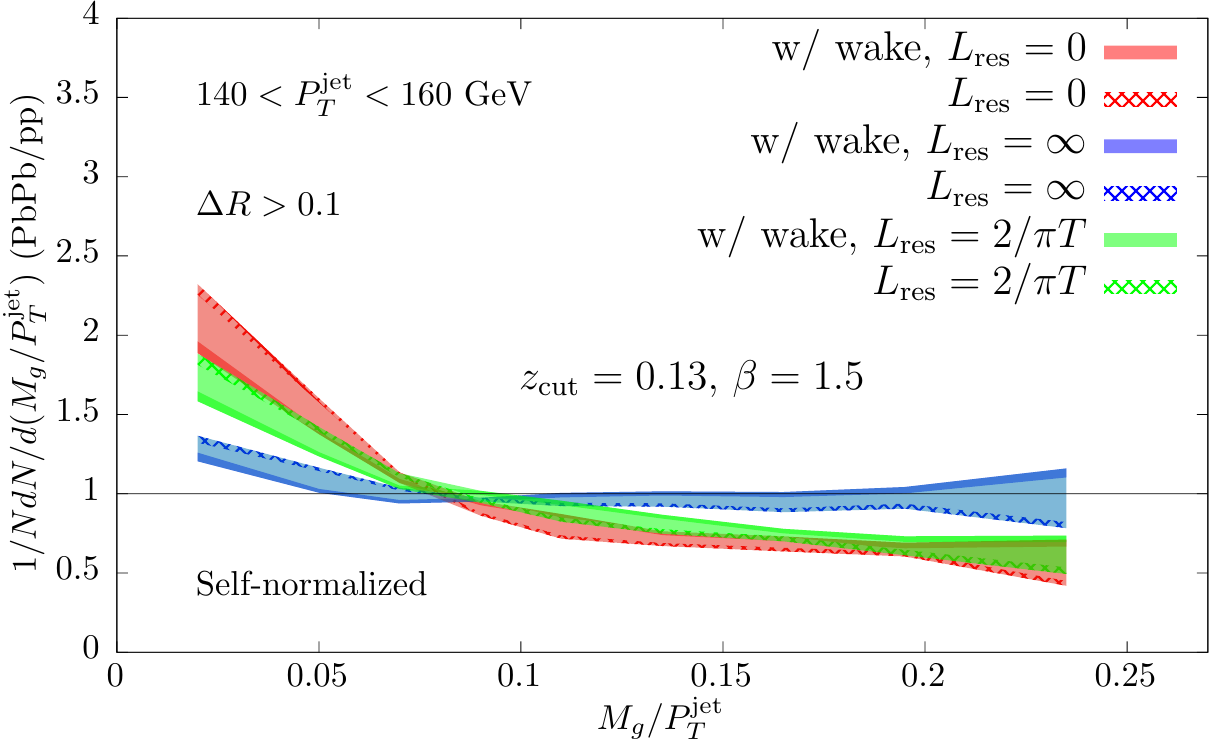}
\caption{\label{Fig:mgLresfinite} Results from the hybrid model calculations
of the ratio between the self-normalized $M_g/p_T^{\rm jet}$ distributions in PbPb and pp collisions at $\sqrt{s}=5.02$ ATeV for 0-5\% centrality class, for $R=0.4$ jets, obtained using the soft-core grooming procedure, and computed for different values of $\lres$. Dashed curves are obtained ignoring the soft particles coming from the wake in the medium, while solid curves do include them. The space in between is shaded for each choice of $\lres$ so as to better expose the relative sensitivity of the observable to the effects of the wake (width of each individual colored shading) and the effects of our choice of $\lres$ (separation between the
different colors).
}
\end{figure}

In Fig~\ref{Fig:mgLresfinite}, we show our hybrid
model calculations for the ratio of the $M_g/p_T^{\rm jet}$ 
distributions in medium to those for jets in vacuum, for media with three values
of $\lres$. The calculations were done using the soft-core grooming procedure and the 
individual distributions are self-normalized.
As for the $z_g$ distribution, the $\lres=2/(\pi  T)$ result is distinct from both the  extreme $\lres$ limits.  And, again as for the $z_g$ distribution,  
it is much closer to the 
result obtained for $\lres=0$ when the medium can fully resolve all components of a jet
than to the result obtained for $\lres=\infty$ when the medium cannot resolve any substructure
within a jet.
However, unlike in the case of the $z_g$ distribution, for this observable the effect of soft particles produced by the wake in the medium is comparable to the separation between $\lres=0$ and $\lres=2/\pi T$.
This indicates that using this observable alone to discriminate among realistic values of  $\lres$ would require separately constraining the dynamics of the wake in the medium sufficiently well to 
reliably quantify its
contribution to this observable.
Nevertheless, having two different observables with clear sensitivity to the value of $\lres$  indicates that these types of measurements of groomed jet observables can potentially be used to extract the resolution length of quark-gluon plasma from jet data.

We conclude this Section by noting that there is an experimental disadvantage to using the soft-core grooming procedure unless the experimental analysis can be extended to jets with high $p_T$. With the soft-core choice of Soft Drop grooming parameters, the value of $z_g$ can be as low as $z_g \sim 0.016$ when $\Delta R=0.1$, which translates into 
the transverse momentum of the softer of the two subjets -- that {\it pass} the Soft Drop criterion (\ref{SoftDropCondition}) --
being as low as 2 GeV for the 140 GeV jets used
in the analysis shown in Fig.~\ref{Fig:mgLresfinite}. This fact may make it extremely challenging 
to perform a sensible measurement without including the contamination of the large fluctuating background. A clear solution would be to move to much higher $p_T$ jets. In Appendix~\ref{sec:gmasspt}, we study the jet-$p_T$ dependence of the groomed jet mass observable and show that, in the range $140<p_T<300$ GeV, this observable is almost independent of jet-$p_T$. We therefore expect that similar conclusions can be drawn from the analysis of jets with momentum $\sim$ 1 TeV, for which the softer subjet with $z_g\sim 0.016$ would 
have $16$ GeV in transverse momentum.

\section{Discussion, including a look at present data and at the road ahead}
\label{sec:conc}

One important aspect of understanding the interaction of QCD jets with the 
QGP formed in heavy ion collisions is understanding how different constituents of a jet 
interact with the medium through which they pass. 
As a consequence in part of our limited knowledge about the space-time structure of jet fragmentation, several different model assumptions 
can be found in the literature. In some models, the highly virtual 
energetic colored excitation that evolves and propagates as a jet
behaves in the medium as if it were a single featureless energetic colored probe
in the medium, in essence as if it were a single hard parton, meaning
that the interactions of the jet depend only on its energy and color and are
independent of its inner structure.
In other models, the medium interacts with jet constituents which are assumed to be formed before the jet interacts with the medium interaction. 
Finally, in a third class of models some fragments of the jet form within the medium 
and start to interact with the medium independently from one another according to some criterion.  
While it could be possible to discriminate between these different assumptions by a systematic comparison between many models and many sets of data, it is at least as 
desirable to identify observables that are directly sensitive to these dynamics. 
In this paper we have investigated a set of groomed jet observables and
shown how several of them are sensitive to the substructure within jets if, that is, the
QGP medium has sufficient resolving power.

The question of whether the medium interacts with a jet as if it were a single
featureless object characterized only by its energy and color or whether it
interacts with constituents within a jet is a question about a property of the 
medium~\cite{CasalderreySolana:2012ef} known as the resolution length $\lres$;
it is not a question about the structure of the jets.  
The parameter $\lres$ characterizes the resolving power of the medium vis-a-vis highly energetic colored probes in much the same way that the screening length characterizes how the
medium ``sees'' heavy (nearly) static colored probes.  The screening length can be thought
of as the minimal distance by which two nearby static color charges must be separated
such that there is enough medium between them that 
they are independent of each other (in the sense that they are not bound to each other.)
The resolution length $\lres$ that we can use jet observables to learn 
something about is the minimal distance by which two nearby colored charges
in a jet must be separated such that there is enough medium 
between them that they engage with the plasma, and lose energy, independently.
The limit $\lres=\infty$ realizes the assumption that 
the medium is totally incapable of resolving any substructure
within a jet, meaning that
the interaction of jets with the medium is completely independent of jet structure. 
As $\lres$ decreases, an increasing fraction of the jet components 
will separate from each other sufficiently that at some point during
their evolution while they are within the medium they will begin to
interact with the medium independently of one another. 
In the limit $\lres=0$, all partons within a jet shower interact with the medium
independently of one another from the moment that each of them forms in
a splitting within the evolving shower.

In the hybrid model calculations that we have presented, we have assumed that the jet forms 
while traversing the medium as in the third class of models described in the first paragraph 
of this Section. Upon making this assumption, we have found
that the totally resolved $\lres=0$ limit  and the totally unresolved $\lres=\infty$ limit are clearly distinguishable from one another, since they provide quite 
distinct predictions for two groomed observables that we have identified: the $\Delta R$-dependent $z_g$-distributions of jets with different angular separations groomed using the flat Soft Drop procedure (Section~\ref{sec:zgdist}); and the 
groomed jet mass distribution obtained using the soft-core  grooming procedure (Section~\ref{sec:gmass}).  
We have also seen that the number of soft-drop splittings, $n_{\rm SD}$,
shows some sensitivity to $\lres$ (Section~\ref{sec:nsd}). And, along the way, we have found that looking
at how the the Lund plane distribution that characterizes the pair of subjets that first
satisfies the Soft Drop condition during the grooming procedure is particularly helpful
in understanding the physical phenomena behind these observables (Section~\ref{sec:lundplane}). 
All of these groomed jet observables  have greater utility in this regard than
the ungroomed charged jet mass (Section~\ref{sec:chmass}) since applying
the Soft Drop grooming procedure reduces the sensitivity of the groomed observables
to particles coming from the wake that the jet leaves in the medium.

The origin of the discriminating power of the groomed jet observables that
we have focused on is the dependence of jet quenching on the activity of the jet.
Loosely speaking, this refers to the number of constituents within a jet; one
of the ways of quantifying it is the number of Soft Drop splittings, $n_{\rm SD}$.
In jet showers whose constituents are resolved by the medium in which
they find themselves,
those jets that are made up of more constituents as seen by the medium
possess more independent sources of energy loss, and as a consequence they lose more energy than jets with less activity. 
Since, as we have explicitly checked in Appendix~\ref{sec:correlation}, 
jet activity is correlated with the angular separation $\Delta R$ between 
the first pair of subjets that pass the Soft Drop condition (\ref{SoftDropCondition}), 
wide jet structures lose more energy than narrow jets. 
Because the jets that remain in the sample of jets after quenching 
are those that have lost the least energy, if the medium has a short enough $\lres$ that
it can resolve jet substructure the consequence will be that the sample of
jets after quenching will be biased toward being narrower (smaller $\Delta R$),
having a lower groomed jet mass (smaller $M_g$) and
having less activity (smaller $n_{\rm SD}$).
In contrast, if the medium cannot resolve any structure within jets then
jet quenching will introduce none of these biases.
Note that this mechanism does not discriminate 
when jets  fragments are formed, provided they are formed in the medium. 
We have explicitly tested that a variation (an unphysical variation) 
of our model in which all jet fragments are produced at a formation time $\tau_f=0$
that is before they pass through any medium leads to comparable
results for these observables (see Appendix~\ref{sec:formationtime}).
This analysis 
 also demonstrates that formation-time effects do not contribute significantly to the 
 results that we have presented. 
 This means that although the observables that we have considered
 can be used to learn about the resolution length $\lres$ of QGP 
 they are not sensitive to the space-time structure of the jet shower,
 making them of limited use for tomography.
 It would be interesting to design new observables that are more 
 sensitive to the space-time history of the jet shower within the medium.

 Some of the observables that we have analyzed have already been already measured, both at LHC and RHIC energies, by the CMS \cite{Sirunyan:2017bsd}, STAR \cite{Kauder:2017cvz} and ALICE \cite{Acharya:2019djg} collaborations. 
 However, none of the measurements available to us today
 have been fully unfolded. This means that a direct comparison between our results and extant measurements is not possible. However, very recently the ALICE collaboration has 
 indirectly compared the predictions of our 
 hybrid model calculations for a medium that can fully resolve
 the partonic substructure of a jet, $\lres=0$,
  for the $z_g$-distribution (distribution of the momentum sharing between the 
  first pair of subjets that passes the Soft Drop condition) with their measured data on 
  this observable
  by embedding our Monte Carlo 
  results into their experimental analysis, smearing the results of our calculations such that they
  can then be compared with their measured data.
  The ratio of the $z_g$-distribution in PbPb and pp collisions, 
  as measured by 
  both CMS and ALICE in data that has not been unfolded,
  show a clear $z_g$-dependence. This is apparently unlike our results, 
  as presented in Fig.~\ref{Fig:zgdelR}, where we find that this ratio depends little on $z_g$.
  After our theoretical predictions for the case of a medium with $\lres=0$ that 
  can fully resolve jet constituents are embedded in the experimental 
  analysis and in this way smeared,
  they also show 
 a significant $z_g$-dependence, compatible with ALICE measurements; see the plots in Ref.~\cite{Acharya:2019djg}. 
 This observation also indicates that the $z_g$-dependence in the PbPb/pp ratio
 of $z_g$ distributions exhibited by the CMS measurements may also
 be significantly corrected after the data are unfolded or the predictions are smeared.  For this reason, in this paper we have refrained from  making direct comparisons with experimental
 data. 
 
 In our model, hard medium induced splittings that propagate out of the medium are absent. The presence of large-angle medium-induced radiation and consequently large-angle medium-induced
 components of jets 
is a characteristic of the perturbative treatment of jet energy loss, but they are absent at strong coupling. 
 If these hard modes (hard relative to the thermal scale that characterizes the medium) leave the collision zone, they lead to a modification of the $z_g$ distribution. Since this radiation is dominantly soft relative to the jet momentum, such an effect leads to an enhancement of the small-$z_g$ region in the momentum-sharing fraction distribution. The direct comparison of calculations based on this physics~\cite{Chien:2016led,Mehtar-Tani:2016aco,Chang:2017gkt}
with non-unfolded CMS data  \cite{Sirunyan:2017bsd} showed a reasonably good agreement. 
As we have stressed when comparing our results in Fig.~\ref{Fig:zgdelR} and the folded results produced by ALICE, 
such comparisons may be misleading.  
When our results are smeared in order to compare them to non-unfolded ALICE data,
they change in qualitative ways and we can expect comparable changes
also if calculational results from Refs.~\cite{Chien:2016led,Mehtar-Tani:2016aco,Chang:2017gkt}
were smeared as required for comparison to non-unfolded CMS data.
In addition, the perturbative mechanism that we have just described 
would imply some shift in the $n_{\rm SD}$-distribution toward larger $n_{\rm SD}$, 
since the mechanism incorporates 
additional splittings in PbPb jet showers. In contrast, 
in our findings  for resolved showers we see a reduction in $n_{\rm SD}$ as compared to 
that in pp collisions, see Fig.~\ref{Fig:nsd}. It also worth pointing out that ALICE measurements of the $n_{\rm SD}$ favor a shift toward lower $n_{\rm SD}$, namely passage through the medium
yielding a reduction 
in the number of splittings and $n_{\rm SD}$,
in qualitative
agreement with the results of our calculations. 
This seems to disfavor any scenario that increases the number of splittings. 
Furthermore, in perturbation theory it is also expected that, unless the formation time of the medium-induced gluon is longer than the medium length, 
  medium-induced radiation should suffer a quick degradation as a consequence of rescattering with the medium constituents \cite{Blaizot:2013hx}, which should
  significantly reduce the increase in $n_{\rm SD}$ 
  relative to a naive analysis that does not include these secondary interactions. Finally, 
  based on a separation of scales argument, 
  we have also not included any medium modification of the high-virtuality stage of the evolution, as done in \cite{Cao:2017zih}, which could also alter the $z_g$-distribution.

While firm conclusions must await quantitative 
comparison with unfolded data, let us return to the  comparison in Ref.~\cite{Acharya:2019djg} 
between ALICE results and our smeared fully resolved jet simulations (medium with $\lres=0$) 
for the purpose of extracting some tentative lessons from those measurements. 
 As in our calculations for a medium with $\lres=0$, and consistent with our qualitative 
 expectations for a medium that is capable of resolving substructure within jet showers,
 ALICE has observed a significant change in the normalization of the 
 $\Delta R$-dependent $z_g$ distribution depending on the constraint applied to $\Delta R$. 
 As in our analysis, the total fraction  of narrow jets (jets with two subjets that pass the Soft Drop condition that are separated  by $\Delta R<0.1$) is larger in the ensemble of PbPb jets
 than in pp; conversely, the total fraction of wide jets (jets with two subjets separated by $\Delta R>0.2$)
 is smaller. 
 Furthermore, the ratio of the $z_g$ distribution in PbPb to that in pp collisions 
 extracted from 
 these measurements are quite similar to what the ALICE collaboration obtains
 from smearing the results of our $\lres=0$ calculations of fully resolved jets.
 That said, after smearing our $\lres=0$ calculations somewhat over predict the deviation
 of the ratio of $z_g$ distributions from one. That is, 
 the results for the PbPb/pp ratio of $z_g$ distributions
 for $\Delta R<0.1$ (for $\Delta R>0.2$) from our $\lres=0$ calculations after smearing lie 
 somewhat above (below) the measured ratios, in both cases deviating farther from one.
  In light of this comparison, it is tempting to conclude that these measurements indicate that the resolution length
 $\lres$ is finite, greater than 0 but certainly not infinite,
 as we expect on general grounds anyway.  It would be interesting to compare
 the results that we have obtained for a medium with $\lres=2/(\pi T)$, shown
 in Section ~\ref{sec:constrainingLres}, after embedding them in the experimental
 analysis and smearing them, to the measured data.

 It will be very exciting to make a systematic comparison with data 
  to test how tightly the measured results constrain the resolution
  length of quark-gluon plasma $\lres$.  Doing so either requires unfolded
  data or smeared calculations with varying values
  of $\lres$.
 We can already see today, however, that the possibility that QGP may behave
 as a medium with $\lres=\infty$, meaning that it resolves no jet substructure whatsoever,
 is quite clearly disfavored by experimental data.
 Our calculations in
 Section~\ref{sec:zgdist} demonstrate clearly that if 
 the medium were to behave as if $\lres=\infty$, there would be no separation between
 the PbPb/pp ratio of $z_g$ distributions with $\Delta R<0.1$ and those with $\Delta R>0.2$.
 This is an example of a qualitative conclusion from our analyses that we expect not
 to depend on any details of the hybrid model.
 And, in contrast, in the experimental data~\cite{Acharya:2019djg} there is a clear separation.
 Further evidence for the conclusion that the resolution length of QGP is short enough that
 the QGP produced in heavy ion collisions {\it can} resolve at least some jet substructure
 comes from the comparison between $R_{\rm AA}$ for jets and $R_{\rm AA}$ for high-$p_T$ hadrons,
 as described in Ref.~\cite{Casalderrey-Solana:2018wrw} and Appendix~\ref{sec:raa}.
We expect that as the precision of experimental measurements of jet substructure observables and the control over uncertainties in their calculation improves, it will become possible 
to constrain the value of the resolution length of QGP, in addition to seeing how the substructure of jets is modified via their passage through it. The road ahead toward these goals seems clear.

We would also like to compare our model results with other model analyses of 
groomed jet observables. 
A particularly salient example is JEWEL~\cite{Zapp:2012ak}, a Monte Carlo event generator
based on a perturbative treatment of jet-medium interactions, including 
medium-induced gluon radiation as well as hard recoils due to the interaction of energetic partons with medium constituents \cite{KunnawalkamElayavalli:2017hxo}. This model has been very successful in predicting 
the $z_g$ distributions measured
by both in CMS and ALICE. Unlike in our model calculations
in which effects coming from the wake that the jet leaves in the plasma are negligible,
in the JEWEL calculations the recoiling particles from the medium are relatively 
hard and so make more of a contribution to groomed jet observables,
and hence are crucial to describing experimental measurements of these
observables~\cite{Milhano:2017nzm}. 
However, at least with current subtraction methods, this model seems unable
to describe the charged jet mass distributions~\cite{Acharya:2017goa}. To the best of our knowledge, our work provides the first simultaneous description of both the
charged jet mass distribution and, once smearing effects are properly addressed, the $z_g$-distribution.

{\bf Note added:} During the completion of this work, Ref.~\cite{Caucal:2019uvr} appeared. These authors consider some of the same observables that we do (in particular, $z_g$ for different 
constraints on $\Delta R$ and $n_{\rm SD}$), and reach similar conclusions for how these observables are modified in PbPb collisions, working within a rather different model than the one that we have developed.  This provides evidence for the robustness of our 
conclusions and, in particular, highlights that in order to explain these 
phenomena it is necessary that jets containing more constituents tend to lose more energy and tend to contain subjets at larger values of $\Delta R$.

\acknowledgments

We are grateful for helpful conversations with Harry Andrews, Chiara Bianchin, Jasmine Brewer, Yi Chen, Leticia Cunqueiro, Matteo Cacciari, Zachary Hulcher, Peter Jacobs, Yen-Jie Lee, Chris McGinn, Tan Luo, Yacine Mehtar-Tani, Konrad Tywoniuk, Marta Verweij, Korinna Zapp and Nima Zardoshti.  

The work of JCS is supported by grants SGR-2017-754, FPA2016-76005-C2-1-P and MDM-2014-0367.
The work of JGM is supported by Funda\c  c\~ao para a Ci\^encia e a Tecnologia (Portugal) under project CERN/FIS-PAR/0022/2017.
The work of DP is supported by a grant from the Trond Mohn Foundation (project no.~BFS2018REK01) and by US NSF grant ACI-1550300.
The work of KR is supported by the U.S. Department
of Energy under grant Contract Number DE-SC0011090.
JGM and KR gratefully acknowledge the hospitality of the CERN theory group.

\appendix


\section{Background subtraction with medium response}
\label{sec:negatives}

As explained in Section~\ref{sec:model}, 
conservation of momentum implies that the momentum lost by a jet must end up carried
by the QGP fluid; the jet must create a wake of some sort in the fluid, a wake that carries
the momentum lost by the jet.  Because of the presence of some region of fluid that is
boosted in the jet direction, after hadronization there must be an excess (depletion) of
soft particles coming from the medium moving in the jet direction (in the opposite direction
relative to the jet) relative to how the fluid would have hadronized in the absence of any wake.
Since soft particle production from the hydrodynamized wake in the fluid that carries the momentum
lost by the jet  is correlated with the jet direction, 
when jets are later reconstructed in experimental analyses, of necessity some these soft particles
end up reconstructed as a part of the jet. Hence, these
medium-response contributions contribute to any jet observable and should be included in the definition of a given jet signal. In this Appendix we describe how we have treated this correlated-medium contribution in this work.

In a full event simulation, the treatment of this particular soft jet component would be straightforward: it would simply amount to applying the experimental analyses to the Monte Carlo events. However, we do not have a full event simulation; as explained in  Section~\ref{sec:model}, our implementation of how the medium reacts to the jet is based on analyzing the modification of particle production with respect to the undisturbed QGP fluid. As a consequence, the expression for the modification of the mean number of particles, Eq.~(\ref{onebody}), can become negative; we explain in Section~\ref{sec:model} that this is not unphysical, it just reflects the fact that the boosted medium, whose momentum is correlated with that off the jet, produces less particles in certain directions of phase space than what an unperturbed droplet of QGP would produce. 
Our treatment of these ``negative particles'' assumes that within the area of a reconstructed jet, many soft particles from an uncorrelated underlying event are collected, such that there are many particles within a given range of momentum and angle, some of which can be cancelled by the 
negative particles. (In our first paper discussing the effects of medium response~\cite{Casalderrey-Solana:2016jvj}, we have benchmarked our calculations of
the particles coming from the wake in the plasma 
by simulating a full thermal background, adding the ``positive particles'' from Eq.~(\ref{onebody})
and removing particles from the thermal background corresponding to the ``negative particles''
coming from Eq.~(\ref{onebody}), and then applying the actual experimental background subtraction procedures used in the experimental measurements of an observable of interest.
Having validated doing so, we now work directly from Eq.~(\ref{onebody}) without first
simulating a full thermal background.)

With the approximation we have just mentioned, there are two possible ways in which 
soft particles coming from the medium carrying momenta that are correlated with
the jet direction can affect jet measurements: first, 
additional particles produced within the jet area can be reconstructed as
a part of the jet,
increasing the momentum of the jet; second, 
a particle (or particles) that would have been reconstructed as a part of the jet in the absence of
any wake in the medium is in fact not produced from the boosted medium, decreasing
the momentum of the jet.  The effect of this missing particle is then identical
to treating the negative particles as having negative four-momenta.

In our previous work \cite{Casalderrey-Solana:2018wrw}, we use this analogy to subtract from the jet the net momentum of all the negative particles that fell within the jet area. 
In this work, since we want to assess the effect of these particles after the sequential declustering of the Soft Drop procedure is run we need to modify the prescription, to include the effect of individual particles. For this reason,
we resort to a simplified prescription in which we use a variation of recombination algorithms within FastJet that takes into account the presence of the negative momentum particles\footnote{We thank Matteo Cacciari and Tan Luo for providing the code.}. It simply consists of preserving the actual four-momentum of the particle with an added ``status'' tag set to +1 for a positive particle or -1 for a negative one. When two clusters are added, those which are negative are subtracted from the sum. If the combination has negative energy, then its four-momentum is flipped and it is tagged with ``status'' -1. The process is iterated, as usual, until all tracks have been clustered into jets. We have checked that at least for coarse quantities, such as total jet momentum, this procedure gives the same result as the subtraction of the net momentum of negative particles, as  done in  Ref.~\cite{Casalderrey-Solana:2018wrw}.

\section{Jet and hadron suppression for unresolved and fully resolved jets}
\label{sec:raa}

\begin{figure}
\centering 
\includegraphics[width=.7\textwidth]{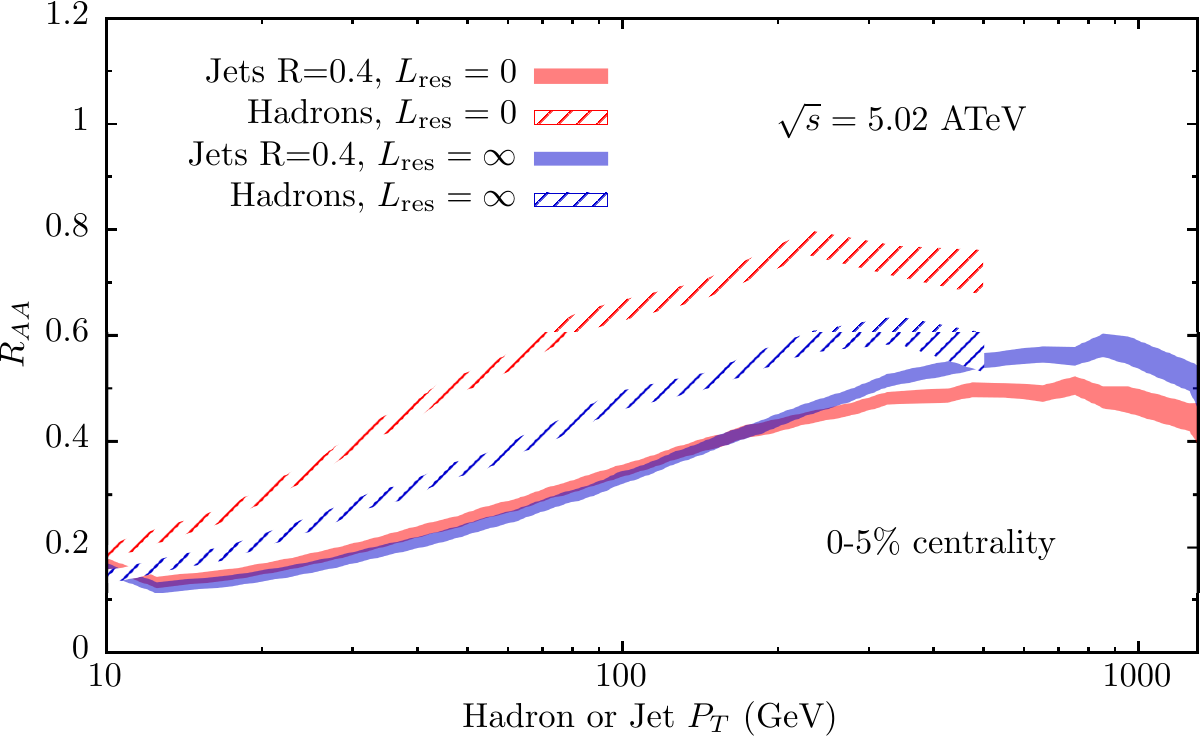}
\caption{\label{Fig:raa} Comparison of our hybrid model calculations of 
$\hraa$ and $\jraa$ for media with $\lres=0$ and $\lres=\infty$.
The results correspond to $0-5\%$ central PbPb collisions with $\sqrt{s}=5.02$ ATeV.}
\end{figure}

In this Appendix we describe how we have fixed the model parameter $\aSC$ for the case
of a medium with $\lres=\infty$ that cannot resolve any structure within jets. Since we want
to use this calculation to 
 determine the sensitivity of jet substructure to resolution effects, the value of $\aSC$ is fixed by demanding that jet samples with $\lres=0$ and with $\lres=\infty$ are produced at similar rates in heavy ion collisions. Since the value of $\aSC$ for the case of a medium with $\lres=0$ 
 that can fully resolve all structure within jets has been accurately determined via a global analysis of jet and hadron suppression data~\cite{Casalderrey-Solana:2018wrw}, we use this calculation for fully resolved jets as the reference. 
 Setting the temperature below which no further parton energy loss occurs to 
 $T_c=145$~MeV, as in the $\lres=0$ calculation, we demand that  the $\jraa$ at a reference value of jet $p_T\sim 100$ GeV is the same (and with the same spread) in both the $\lres=0$ and $\lres=\infty$ computation. This criterion yields a value of $0.5<\aSC<0.52$ for the $\lres=\infty$ limit, 
 about $25\%$ larger than the values obtained in the fully resolved $\lres=0$  limit. 

It is certainly to be expected 
that the value of $\aSC$ needed to achieve the same jet quenching 
is greater for $\lres=\infty$ than for $\lres=0$ since when $\lres=\infty$ and jets are
completely unresolved, each jet behaves as a single object losing energy whereas
when $\lres=0$ and jets are fully resolved, every parton in each jet shower loses energy
independently.
As shown in Fig.~\ref{Fig:raa}, while the jet spectrum is only matched at one reference value
of the jet $p_T$, here 150 GeV, the two calculations give comparable $\jraa$ in most of the range of transverse momenta of interest. Some deviations are observed in the high-$p_T$ region of $\jraa$. 
This is because the number of partons in a jet shower increases slowly with the jet $p_T$, meaning
that in the case where the medium has $\lres=0$ the jet suppression is greater than 
in the case where the medium has $\lres=\infty$ at values of $p_T$ that are far above the 
reference value of $p_T$.

The larger value of $\aSC$ for the medium that cannot resolve any jet substructure 
has important consequences for the hadron suppression pattern, 
since $\hraa$ is sensitive to the suppression of the leading parton 
within each jet. In Fig.~\ref{Fig:raa}
we also show the results of 
our hybrid model calculations for media with $\lres=0$ and $\lres=\infty$ for $\hraa$.
As shown in the plot, while the $\jraa$ are comparable in the two cases,
the results of the two $\hraa$ calculations are significantly different, with $\hraa$  in the medium that cannot resolve any jet structure being smaller than in the $\lres=0$ case. 
This is, once again, a consequence of the way the medium interacts with the jet constituents. In the case of a medium with $\lres=0$ that can resolve all partons within a jet, 
jet suppression results from quenching multiple jet components. Therefore, for a 
fixed amount of total jet quenching, the leading hadron 
needs to be relatively less quenched. On the contrary, the stronger quenching needed
to obtain the same $\jraa$  in 
the $\lres=\infty$ case where jets are completely unresolved
 implies that the leading parton, and as a consequence the hadron spectrum, is 
 more suppressed than in the totally resolved limit. 

The clear separation in the results for the hadron spectra in the $\lres=\infty$ medium relative to that in the $\lres=0$ medium seen in Fig.~\ref{Fig:raa},
together with the fact that the global analysis of Ref.~\cite{Casalderrey-Solana:2018wrw} shows that the $\jraa$ and $\hraa$ obtained with $\lres=0$ describe all extant 
jet and hadron suppression data from the LHC rather well,  indicate that our $\lres=\infty$ calculation {\it cannot} describe 
the measured values of $\jraa$ and $\hraa$ in LHC data simultaneously.
For this reason, for the analysis of this paper we have not attempted to do a global fit for $\lres=\infty$, instead simply choosing $\aSC$ so as to fix $\jraa$ as we have described in this Appendix. 

We conclude from this discussion that the measured values of $\jraa$ and $\hraa$ in LHC
data disfavor quenching by a medium with $\lres=\infty$.  This is the same
conclusion that we reach in Section~\ref{sec:conc} via consideration of groomed jet observables.

\section{Correlation between $\Delta R$, $n_{\rm SD}$ and $z_g$ 
}
\label{sec:correlation}

In this Appendix we collect a number of facts concerning correlations among 
different Soft Drop  observables for jets that shower in vacuum, 
as modelled by PYTHIA, facts that are important to understanding the systematics 
of jet quenching in our hybrid model calculation.

\begin{figure}
\centering 
\includegraphics[width=.7\textwidth]{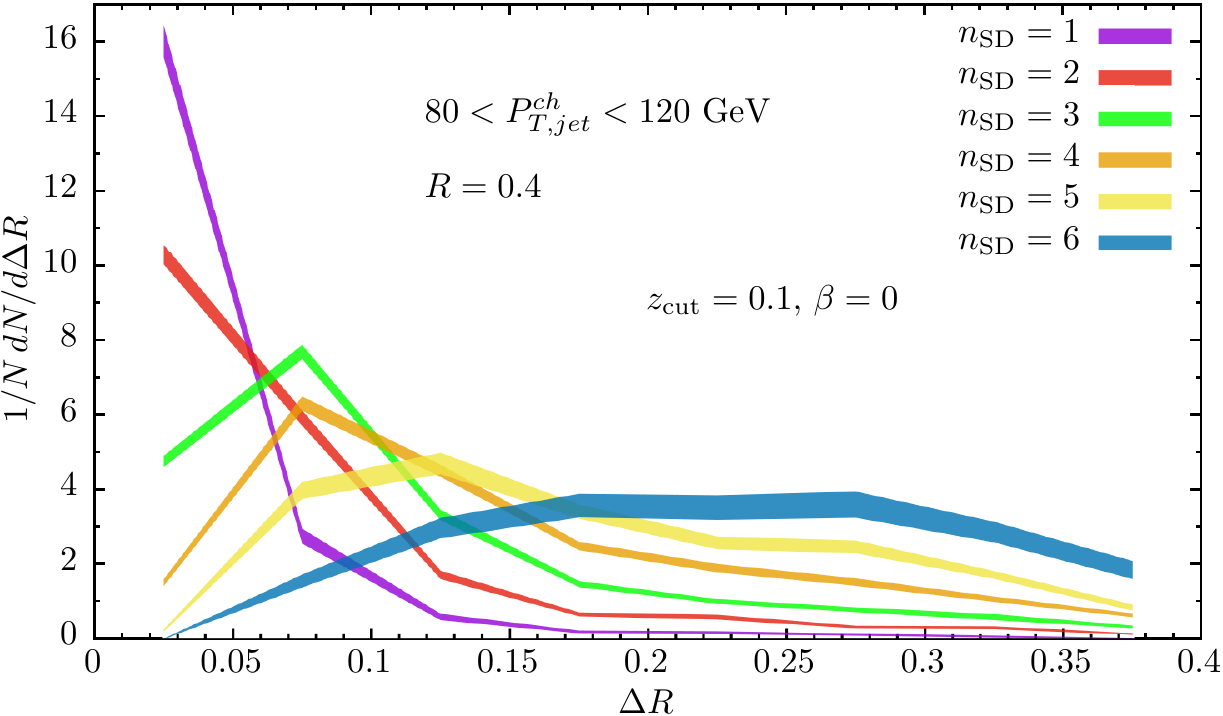}
\caption{\label{Fig:rgvsnsd} The distributions
of the angular separation $\Delta $ between the two subjets identified via 
the first Soft Drop splitting satisfying (\ref{SoftDropCondition})  for jets in vacuum
with different total numbers of Soft Drop splittings, or $n_{\rm SD}$.}
\end{figure}

The first aspect we want to highlight is the strong correlation between the angular 
separation $\Delta R$ of the two subjets identified by the first Soft Drop splitting 
and the total number of Soft Drop splittings,
$n_{\rm SD}$, which is a measure of
the particle multiplicity within the jet. The curves shown in Fig.~\ref{Fig:rgvsnsd} have been generated by classifying pp jets in terms of their $n_{\rm SD}$, and then plotting the distribution of the angle $\Delta R$ between the two subjets at the first of those splittings. 
We see that jets with only one $n_{\rm SD}$ have a very narrow angular distribution, while jets with increasing number of Soft Drop splittings become significantly wider. 
This correlation is easy to understand from the DGLAP shower process.
In this shower, the number of splittings is controlled by the separation between the ordering variable and the regulator of the collinear divergence that stops the fragmentation process \cite{Ellis:1991qj}. In an angular ordered shower, this ordering variable is the splitting angle and, therefore,  large initial angle 
leads to large multiplicity. 
Conversely, if the first splitting occurs at a large angle,
the probability that the jet does not split again, which is controlled by the Sudakov factor, is small since the phase space for radiation is large. 
Therefore and since for angular ordered showers the first splitting coincides with the first declustering step of the Cambridge/Aachen algorithm, 
jets with few $n_{\rm SD}$ come preferentially from narrow jets, with initial angle close to the collinear regulator. 
In practice our jet showers, simulated by PYTHIA8 \cite{Sjostrand:2007gs} are 
$p_T$ ordered. Nevertheless, angular ordering is still approximate in those showers, since these two ordering variables coincide as long as the splitting is not very asymmetric in the  momentum sharing fraction.  Regardless, the bottom line that Fig.~\ref{Fig:rgvsnsd} illustrates is 
that vacuum jets that have a larger $n_{\rm SD}$ and hence a larger
particle multiplicity within the jet tend to have the subjets found at the first Soft Drop
splitting separated by a larger $\Delta R$.

\begin{figure}
\centering 
\includegraphics[width=.7\textwidth]{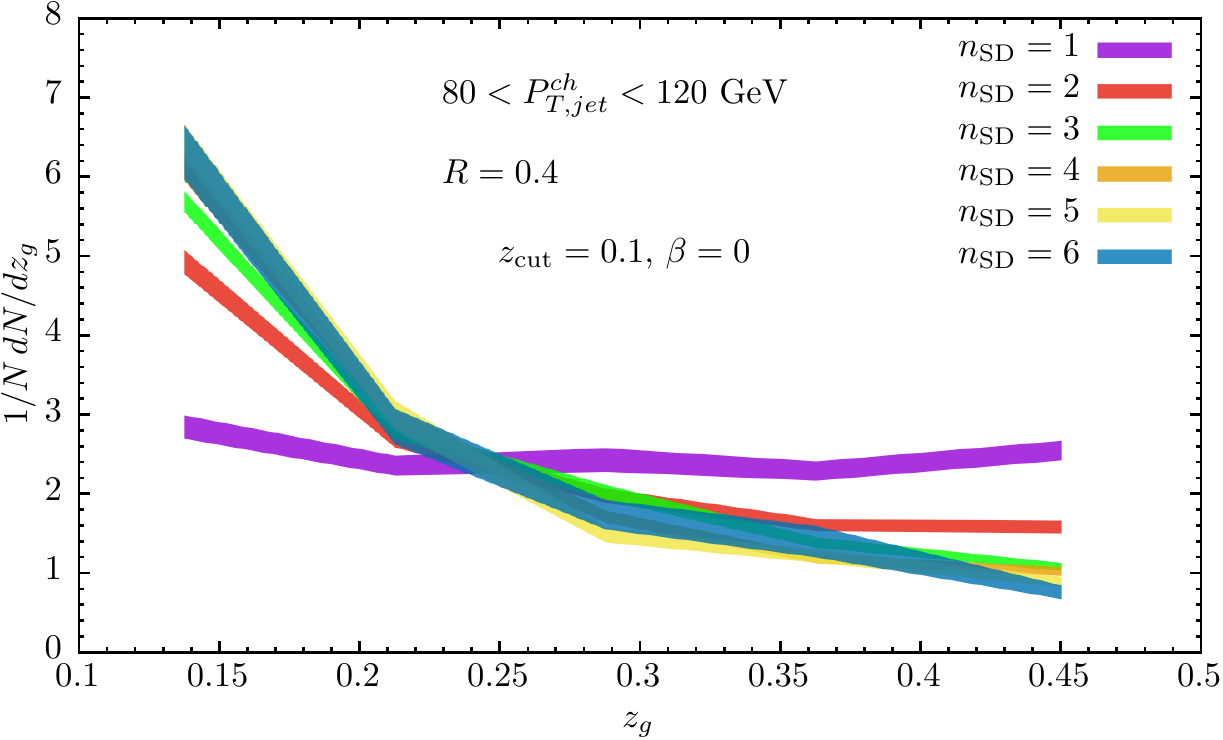}
\caption{\label{Fig:zgvsnsd} The distributions  of the momentum sharing fraction $z_g$ of the first Soft Drop splitting for jets in vacuum that contain different numbers of Soft Drop 
splittings $n_{\rm SD}$.}
\end{figure}

We now look for any correlation between $n_{\rm SD}$ and the $z_g$ of the first Soft Drop splitting. In Fig.~\ref{Fig:zgvsnsd} we 
show the $z_g$ distribution of vacuum jets for various fixed values of $n_{\rm SD}$. One of the main conclusions of this exercise is that as $n_{\rm SD}$ increases, the $z_g$-distribution quickly becomes independent of the number of Soft Drop splittings $n_{\rm SD}$,
and becomes close to the fully inclusive $z_g$-distribution. 
As expected, the  large $n_{\rm SD}$ limit of  this distribution is given by  the Altarelli-Parisi splitting function \cite{Larkoski:2015lea}. It is also curious to see that  for few, $n_{\rm SD}=1,\,2$, the $z_g$ distribution deviates significantly from this asymptotic limit, becoming more and more balanced in $z_g$. For these jets with few constituents,  splittings at small $z_g$ are no-longer dominant, as in the Altarelli-Parisi splitting function. 
The reason for this phenomenon is, once again, the collinear regulator of the shower. Since, in the 
$p_T$-ordered shower of PYTHIA8, splittings happen as long as  $p^2_T/z(1-z)> Q_0^2$    \cite{Sjostrand:2004ef} with $Q_0\sim 1$~GeV the collinear regulator, soft splittings $z\rightarrow 0$ possess a very large phase space for emission. As a consequence, 
for generic fragmentation patterns, the probability of no radiation after the first splitting is very small 
unless the $p_T$ of the splitting is very close to the kinematic limit. 
Since the likelihood of the latter fragmentation patterns are  also suppressed by the probability of no emission between the hard scale and a scale close to the kinematic limit, soft emissions that lead to few (one) $n_{\rm SD}$ are suppressed with respect to the inclusive splitting function. Remarkably, this suppression makes the $z_g$ distribution of jets containing one single $n_{\rm SD}$ approximately independent of $z_g$.

While the two features we have described in this Appendix are properties of vacuum jet showers, they have important implications for understanding groomed jet 
observables in medium.  
First, we conclude that since jets with more constituents have larger
$\Delta R$, see Fig.~\ref{Fig:rgvsnsd},  and since
jets with more constituents lose more energy
the ensemble of jets after quenching will have an $n_{\rm SD}$ 
that is shifted towards lower $n_{\rm SD}$ and a $\Delta R$ distribution
that is shifted towards lower $\Delta R$.
Second, since the $z_g$-distribution of the first Soft Drop splitting very quickly becomes independent of $n_{\rm SD}$ (or the multiplicity) as shown in Fig.~\ref{Fig:zgvsnsd}, 
while energy loss can change the $n_{\rm SD}$ and $\Delta R$ distributions rather 
significantly, the $z_g$ distribution after quenching is much less affected by the quenching process. We have observed both these features in our analysis of groomed observables of quenched jets presented in Section~\ref{sec:gob}.

\section{Sensitivity of substructure observables to the formation time of the splittings}
\label{sec:formationtime}

\begin{figure}
\centering 
\includegraphics[width=.7\textwidth]{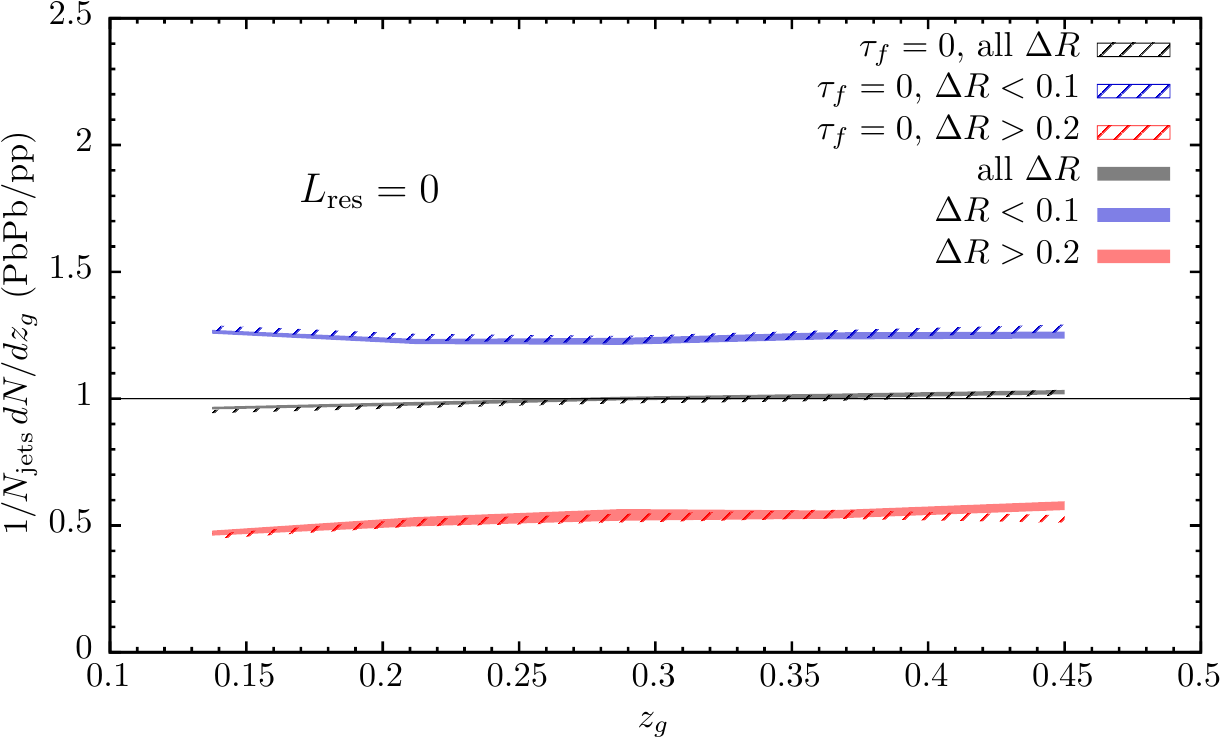}
\caption{\label{Fig:timecheck} Results for the  ratio of the $z_g$ distribution of PbPb jets over that
of pp jets for samples of jets with anti-$k_T$ radius $R=0.4$, with different angular separation $\Delta R$ between the two branches that satisfy the Soft Drop condition (\ref{SoftDropCondition}), using the flat grooming procedure. Each individual curve is normalized to the total number of jets analyzed, $N_{\rm jets}$. We compare results for the completely resolved case $\lres=0$ between two different scenarios: the physical one, with dynamical space-time evolution of the parton shower, depicted with solid bands, versus an unphysical scenario where all on-shell partons are assumed to have been created at the space-time position of the hard scattering, with formation time zero, depicted in dashed bands and labelled as $\tau_f=0$. The solid bands are the same as those in the left panel of Fig.~\ref{Fig:zgdelR}.}
\end{figure}

By comparing the middle panel of Fig.~\ref{Fig:glundInd} to the left panel, we see
that in the case where the medium has $\lres=0$ and can resolve internal structure
within jets (but {\it not} in the right panel where the medium sees each jet as a single
unresolved object) after quenching the peak in the Lund plane distribution is 
at lower values of $\Delta R$ or, better to say, lower
values of $M_g/p_{T, g}$ as can be seen by the lines of constant $\log(1/(M_g/p_{T, g})$ drawn in the Figure. 
This indicates that if the medium can resolve the internal structure of jets, those
with larger values of the groomed jet mass $M_g$ lose more energy.

There is a potential additional physical effect that could also be contributing to the
phenomena represented by the changes to the modulation of the Lund plane distribution
in the middle panel of Fig.~\ref{Fig:glundInd} and the $\Delta R$-dependence of the $z_g$ distribution ratios in the left panel of Fig.~\ref{Fig:zgdelR}.  
In this Appendix, we describe this other physical effect and then show that its contribution
is negligible.

One can express the groomed jet 
mass in terms of the formation time of the splitting $\tau_f$ via 
$\tau_f \simeq 2 \,p_{T,g}/M_g^2$
where we are neglecting rapidity 
and the effects of energy loss in making this estimate. 
That is, in a jet shower in vacuum the wider emissions tend to happen earlier than the narrower ones.
We can ask how much this formation time distinction contributes to the effects
that we are seeing.  To what degree are the jets with larger $\Delta R$ losing more energy because
the splitting responsible for the formation of two subjets separated by a large $\Delta R$
happened earlier? 
We shall show that the answer is: only to a negligible degree.

In Fig.~\ref{Fig:timecheck} we address this issue by comparing our hybrid model calculations 
for the $\Delta R$-dependent $z_g$ distributions for two different setups. 
The first of them, with results depicted in solid bands, is the physically motivated setup used everywhere else in this work (including in Fig.~\ref{Fig:zgdelR})  in which partons are formed sequentially through $1\rightarrow2$ splittings 
after a time $\tau_f =2E/Q^2$, where $E$ and $Q$ refer to the energy and virtuality of the parent parton, respectively. In the second setup, labelled as $\tau_f=0$ and with results 
depicted in dashed bands, 
we make the completely unphysical assumption that all partons (whose virtuality is below 
the infra-red cutoff and will not split further)
were formed with $\tau_f=0$ at the creation point of the hard scattering,
such that they start interacting with the plasma from $t=0$ at $z=0$ and $x_{\perp}^{\rm cre}$. 
We consider this completely unphysical setup because in this case all splittings (including those
that are responsible for producing pairs of subjets with large $\Delta R$ and those
which yield small $\Delta R$) have exactly the same $\tau_f$.\footnote{Given that the total amount of energy loss increases in the $\tau_f=0$ setup compared to the physical one, the values of $\aSC$ have had to be reduced by 5\% such that 
their jet $\jraa$ coincide around $p_T^{\rm jet}\sim 100$ GeV for $R=0.4$ anti-$k_T$ jets, 
in an analogous way to the way in which the values of $\aSC$ for the totally unresolved jets with $\lres=\infty$ where chosen in Appendix~\ref{sec:raa}.}
The conclusion from Fig.~\ref{Fig:timecheck} is rather stark: we have made a rather brutal
change to the formation times, and the $\Delta R$-dependent $z_g$ distributions 
change almost not at all.  This means that the strong $\Delta R$-dependence that
we see in the medium with $\lres=0$ does not originate from differences between
the formation times among different splittings.
Instead, this effect reflects the fact that the medium can resolve the 
structure corresponding to the two subjets, which lose energy independently. Furthermore
 the medium with $\lres=0$ can resolve all subsequent splittings in the jet, and 
 jets with larger $\Delta R$ will on average produce more subsequent splittings, something
 that we confirm explicitly in the previous Appendix.   
 
The exercise that we have done in this Appendix casts some doubt on the
possibility of using the particular jet observables that we consider in this paper for what
has been called tomography.
It is quite clear that different jets lose different amounts of
energy, with those with a large $\Delta R$ from their first Soft Drop splitting losing
more energy, and those with a larger number of
Soft Drop splittings $n_{\rm SD}$ losing more energy.  In contrast, 
making a brutal change to $\tau_f$ and hence a brutal change to how much medium 
the jet partons traverse has very little effect at all.
That is, these groomed observables
possess very little sensitivity to the space-time structure of jet showers, for those splittings which occur within the medium.

\section{Jet momentum dependence of the groomed jet mass}
\label{sec:gmasspt}

\begin{figure}
\centering 
\includegraphics[width=0.7\textwidth]{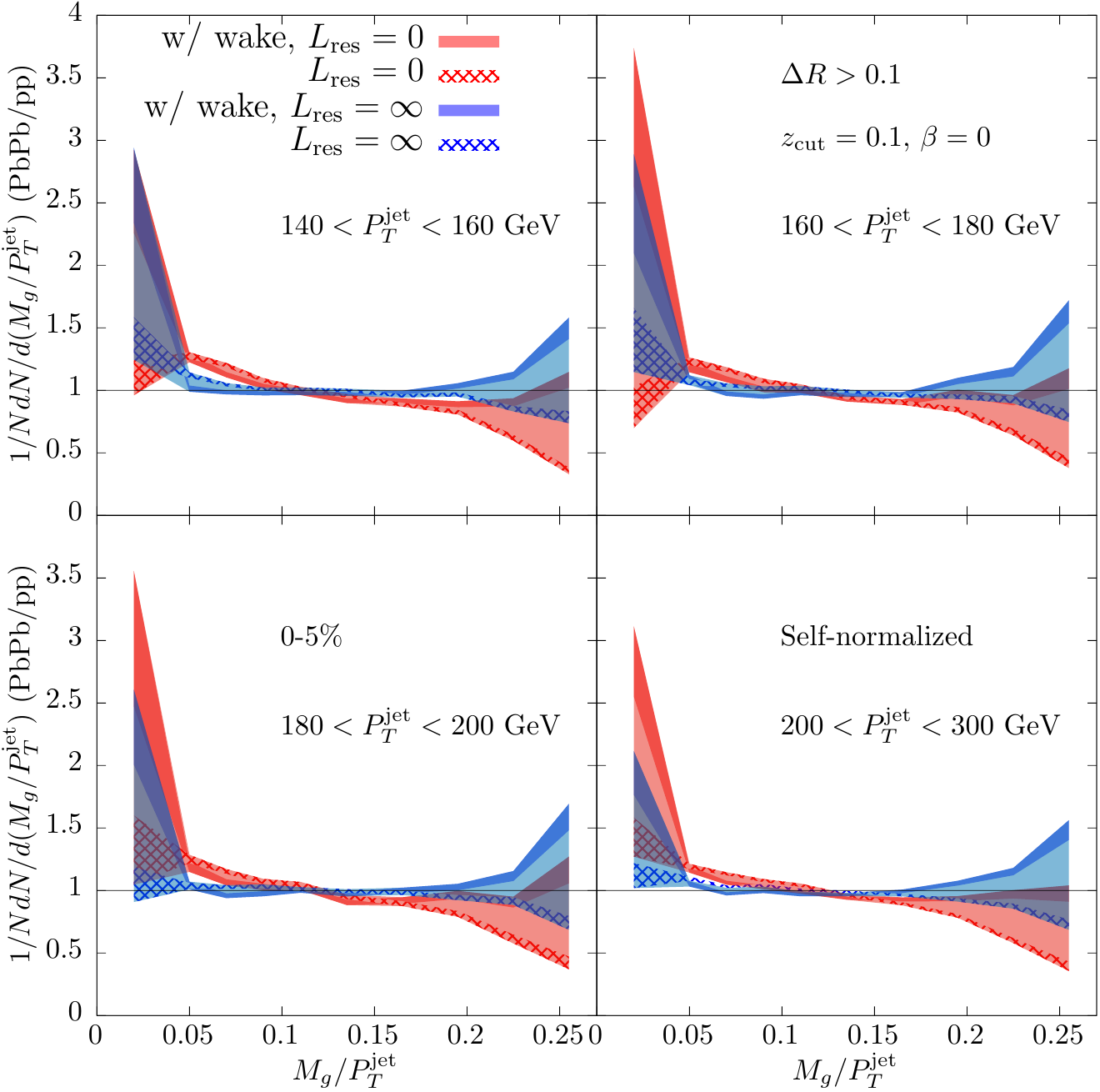}
\caption{\label{Fig:gmassflat} Ratios of the individually 
self-normalized distributions of $M_g/p_T^{\rm jet}$ in PbPb collisions to those in pp collisions, for different ranges of the ungroomed jet momentum $p_T^{\rm jet}$. 
We used the flat grooming procedure, with Soft Drop parameters $z_{\rm cut}=0.1, \, \beta=0$.
Red (blue) curves correspond to model calculations with a medium that has $\lres=0$ ($\lres=\infty$).
Solid curves correspond to the full results which include particles
coming from the wake in the medium; in dashed curves we show results which don't include them.}
\end{figure}

\begin{figure}
\centering 
\includegraphics[width=0.7\textwidth]{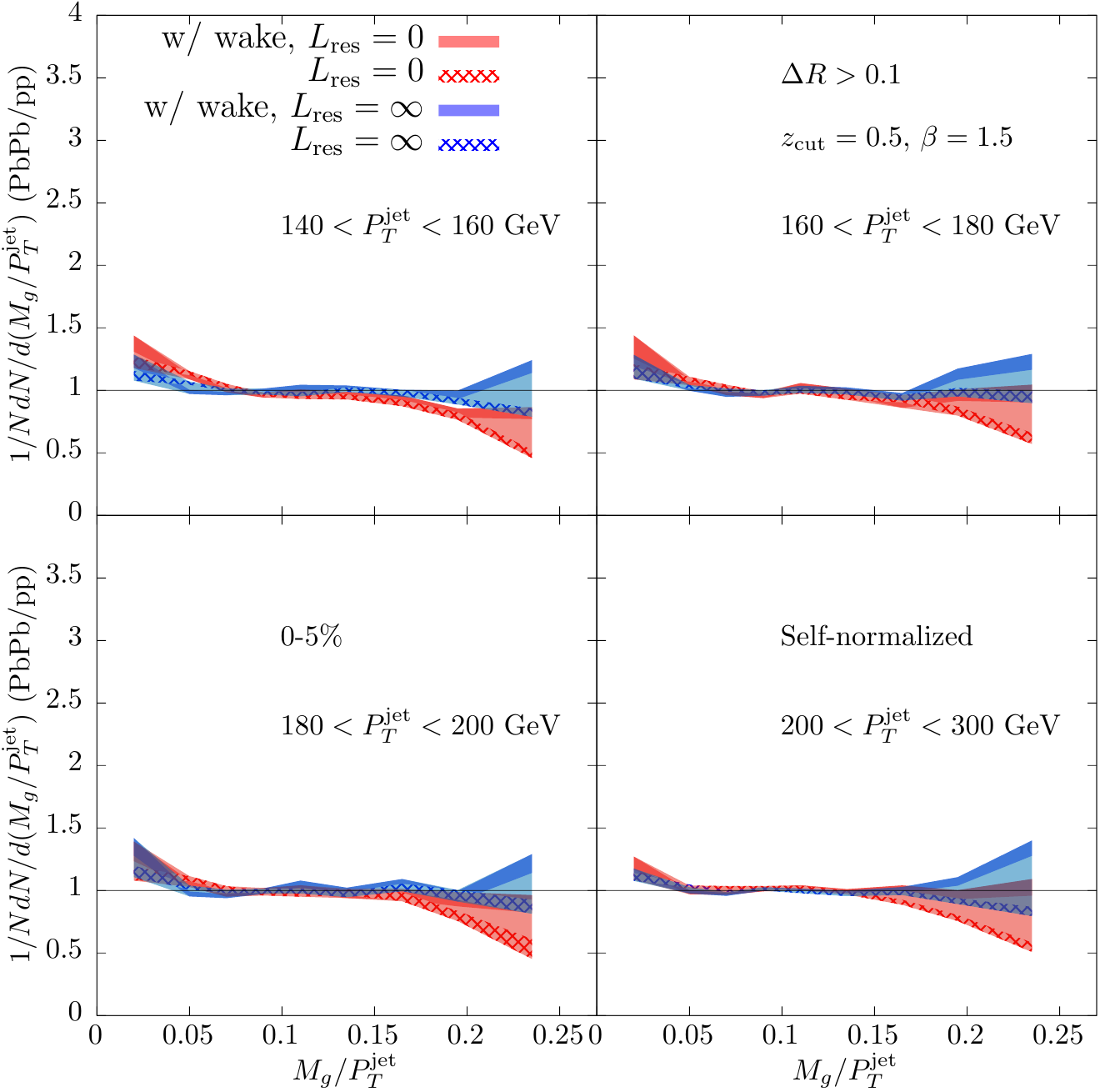}
\caption{\label{Fig:gmasscore} Ratios of the individually self-normalized distributions of $M_g/p_T^{\rm jet}$ in PbPb collisions to those in pp collisions, for different ranges of the un-groomed jet momentum $p_T^{\rm jet}$. We used the core grooming procedure, with Soft Drop parameters $z_{\rm cut}=0.5, \, \beta=1.5$.
Red (blue) curves correspond to model calculations with a medium that has $\lres=0$ ($\lres=\infty$).
Solid curves correspond to the full results which include particles
coming from the wake in the medium; in dashed curves we show results which don't include them.}
\end{figure}

\begin{figure}
\centering 
\includegraphics[width=0.7\textwidth]{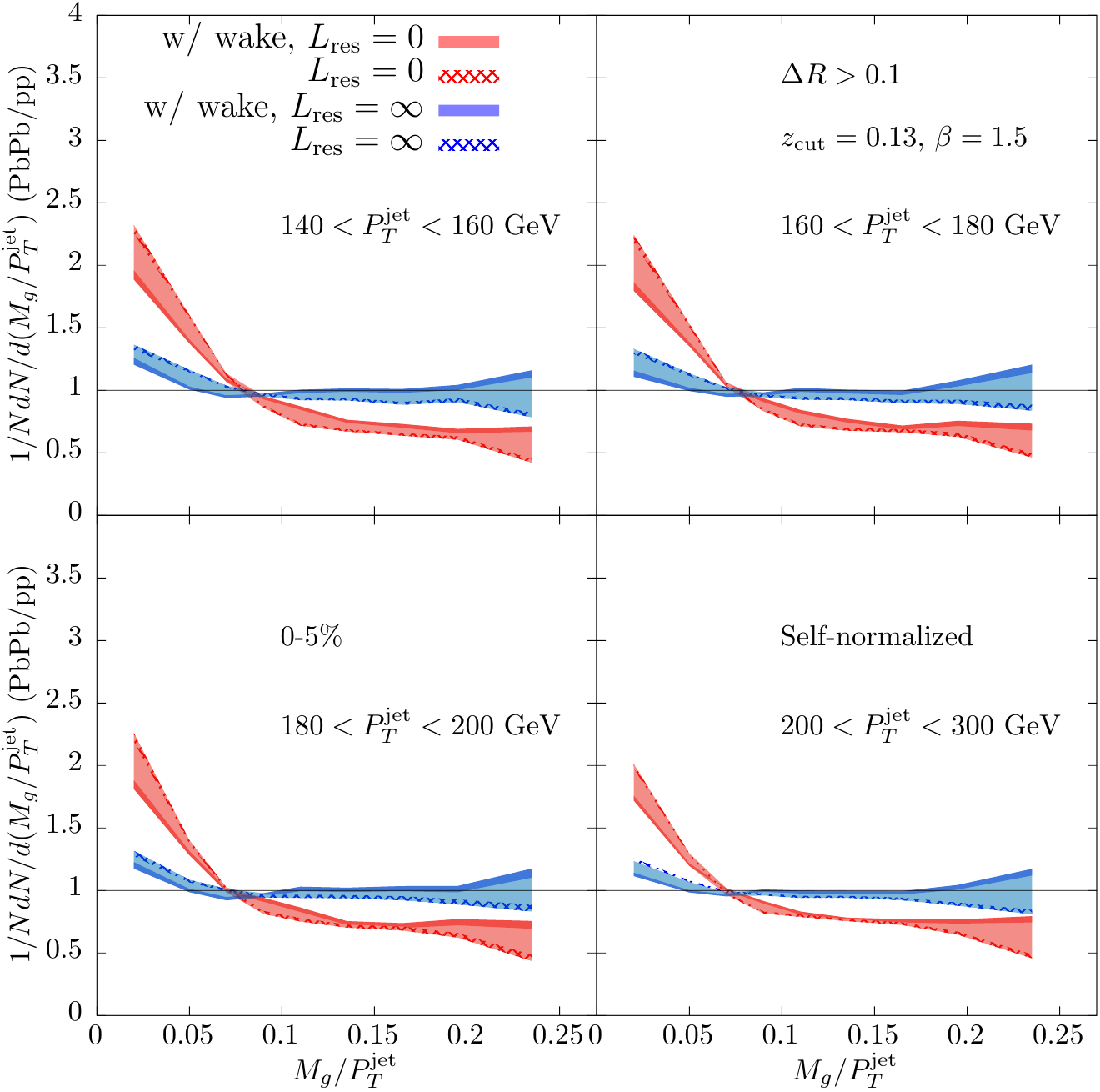}
\caption{\label{Fig:gmasssoftcore} Ratios of the individually self-normalized distributions of $M_g/p_T^{\rm jet}$ in PbPb collisions to those in pp collisions, for different ranges of the ungroomed jet momentum $p_T^{\rm jet}$. We used the soft-core grooming procedure, with Soft Drop parameters $z_{\rm cut}=0.13, \, \beta=1.5$.
Red (blue) curves correspond to model calculations with a medium that has $\lres=0$ ($\lres=\infty$).
Solid curves correspond to the full results which include particles
coming from the wake in the medium; in dashed curves we show results which don't include them.}
\end{figure}

To further connect our analysis with jet measurements at the LHC, in this Appendix we present our results for the ratio of the distribution of the groomed jet mass  divided by the ungroomed jet momentum, $M_g/p_T^{\rm jet}$,  in PbPb collisions to that in pp collisions at $\sqrt{s}=5.02$ ATeV, for the different ranges of the $p_T^{\rm jet}$ for jets with anti-$k_T$ radius $R=0.4$. This has been measured by CMS using  the flat and core grooming procedures~\cite{Sirunyan:2018gct}. Note once again that since CMS data are not unfolded, a direct comparison of our calculation with these measurements is not possible at present. 
After the discussion in Section~\ref{sec:gmass} we only show results for the self-normalised, Sudakov safe distributions. Each Figure corresponds to a different grooming procedure: 
flat grooming in Fig.\ref{Fig:gmassflat}, core grooming in Fig.\ref{Fig:gmasscore} and the 
soft-core grooming that we have introduced, in Fig.\ref{Fig:gmasssoftcore}. 
One can observe that the results are practically independent of $p_T^{\rm jet}$ in the range accessible to current measurements.


\begin{thebibliography}{99}

\bibitem{Adams:2015hiv} 
  D.~Adams {\it et al.},
  Eur.\ Phys.\ J.\ C {\bf 75}, no. 9, 409 (2015)
  [arXiv:1504.00679 [hep-ph]].

\bibitem{Larkoski:2017jix} 
  A.~J.~Larkoski, I.~Moult and B.~Nachman,
  arXiv:1709.04464 [hep-ph].
  
\bibitem{CasalderreySolana:2004qm} 
  J.~Casalderrey-Solana, E.~V.~Shuryak and D.~Teaney,
  J.\ Phys.\ Conf.\ Ser.\  {\bf 27}, 22 (2005)
  [hep-ph/0411315].
  
\bibitem{Neufeld:2008fi} 
  R.~B.~Neufeld, B.~Muller and J.~Ruppert,
  Phys.\ Rev.\ C {\bf 78}, 041901 (2008)
  [arXiv:0802.2254 [hep-ph]].

\bibitem{DEramo:2012uzl} 
  F.~D'Eramo, M.~Lekaveckas, H.~Liu and K.~Rajagopal,
  JHEP {\bf 1305}, 031 (2013)
  [arXiv:1211.1922 [hep-ph]].

\bibitem{Kurkela:2014tla} 
  A.~Kurkela and U.~A.~Wiedemann,
  Phys.\ Lett.\ B {\bf 740}, 172 (2015)
  [arXiv:1407.0293 [hep-ph]].

\bibitem{DEramo:2018eoy} 
  F.~D'Eramo, K.~Rajagopal and Y.~Yin,
  JHEP {\bf 1901}, 172 (2019)
  [arXiv:1808.03250 [hep-ph]].




\bibitem{Milhano:2015mng} 
  J.~G.~Milhano and K.~C.~Zapp,
  Eur.\ Phys.\ J.\ C {\bf 76}, no. 5, 288 (2016)
  [arXiv:1512.08107 [hep-ph]].
  
\bibitem{Rajagopal:2016uip} 
  K.~Rajagopal, A.~V.~Sadofyev and W.~van der Schee,
  Phys.\ Rev.\ Lett.\  {\bf 116}, no. 21, 211603 (2016)
  [arXiv:1602.04187 [nucl-th]].
  
\bibitem{Brewer:2017fqy} 
  J.~Brewer, K.~Rajagopal, A.~Sadofyev and W.~Van Der Schee,
  JHEP {\bf 1802}, 015 (2018)
  [arXiv:1710.03237 [nucl-th]].
  
\bibitem{Casalderrey-Solana:2016jvj} 
  J.~Casalderrey-Solana, D.~Gulhan, G.~Milhano, D.~Pablos and K.~Rajagopal,
  JHEP {\bf 1703}, 135 (2017)
  [arXiv:1609.05842 [hep-ph]].

\bibitem{Larkoski:2014wba} 
  A.~J.~Larkoski, S.~Marzani, G.~Soyez and J.~Thaler,
  JHEP {\bf 1405}, 146 (2014)
  [arXiv:1402.2657 [hep-ph]].
  

\bibitem{Mehtar-Tani:2013pia} 
  Y.~Mehtar-Tani, J.~G.~Milhano and K.~Tywoniuk,
  Int.\ J.\ Mod.\ Phys.\ A {\bf 28}, 1340013 (2013)
  [arXiv:1302.2579 [hep-ph]].

\bibitem{Qin:2015srf} 
  G.~Y.~Qin and X.~N.~Wang,
  Int.\ J.\ Mod.\ Phys.\ E {\bf 24}, no. 11, 1530014 (2015)
  [arXiv:1511.00790 [hep-ph]].




  
\bibitem{Baier:1996kr} 
  R.~Baier, Y.~L.~Dokshitzer, A.~H.~Mueller, S.~Peigne and D.~Schiff,
  Nucl.\ Phys.\ B {\bf 483}, 291 (1997)
  [hep-ph/9607355].
  
\bibitem{Zakharov:1996fv} 
  B.~G.~Zakharov,
  JETP Lett.\  {\bf 63}, 952 (1996)
  [hep-ph/9607440].
  
\bibitem{Baier:1998kq} 
  R.~Baier, Y.~L.~Dokshitzer, A.~H.~Mueller and D.~Schiff,
  Nucl.\ Phys.\ B {\bf 531}, 403 (1998)
  [hep-ph/9804212].
  
\bibitem{Gyulassy:2000er} 
  M.~Gyulassy, P.~Levai and I.~Vitev,
  Nucl.\ Phys.\ B {\bf 594}, 371 (2001)
  [nucl-th/0006010].
  
\bibitem{Wiedemann:2000za} 
  U.~A.~Wiedemann,
  Nucl.\ Phys.\ B {\bf 588}, 303 (2000)
  [hep-ph/0005129].
  
\bibitem{Wang:2001ifa} 
  X.~N.~Wang and X.~F.~Guo,
  Nucl.\ Phys.\ A {\bf 696}, 788 (2001)
  [hep-ph/0102230].
  
\bibitem{Arnold:2002ja} 
  P.~B.~Arnold, G.~D.~Moore and L.~G.~Yaffe,
  JHEP {\bf 0206}, 030 (2002)
  [hep-ph/0204343].
  
\bibitem{Salgado:2003gb} 
  C.~A.~Salgado and U.~A.~Wiedemann,
  Phys.\ Rev.\ D {\bf 68}, 014008 (2003)
  [hep-ph/0302184].
  
\bibitem{Jeon:2003gi} 
  S.~Jeon and G.~D.~Moore,
  Phys.\ Rev.\ C {\bf 71}, 034901 (2005)
  [hep-ph/0309332].
  
  
    
\bibitem{Jacobs:2004qv}
  P.~Jacobs and X.~N.~Wang,
  Prog.\ Part.\ Nucl.\ Phys.\  {\bf 54}, 443 (2005)
  [arXiv:hep-ph/0405125].

\bibitem{Lokhtin:2005px} 
  I.~P.~Lokhtin and A.~M.~Snigirev,
  Eur.\ Phys.\ J.\ C {\bf 45}, 211 (2006)
  [hep-ph/0506189].
  

\bibitem{CasalderreySolana:2007zz}
J.~Casalderrey-Solana, C.~A.~Salgado,
Acta Phys.\ Polon.\  {\bf B38}, 3731-3794 (2007).
[arXiv:0712.3443 [hep-ph]].

\bibitem{Zapp:2008af} 
  K.~Zapp, J.~Stachel and U.~A.~Wiedemann,
  Phys.\ Rev.\ Lett.\  {\bf 103}, 152302 (2009)
  [arXiv:0812.3888 [hep-ph]].

\bibitem{Zapp:2008gi} 
  K.~Zapp, G.~Ingelman, J.~Rathsman, J.~Stachel and U.~A.~Wiedemann,
  Eur.\ Phys.\ J.\ C {\bf 60}, 617 (2009)
  [arXiv:0804.3568 [hep-ph]].
  
\bibitem{Lokhtin:2008xi} 
  I.~P.~Lokhtin, L.~V.~Malinina, S.~V.~Petrushanko, A.~M.~Snigirev, I.~Arsene and K.~Tywoniuk,
  Comput.\ Phys.\ Commun.\  {\bf 180}, 779 (2009)
  [arXiv:0809.2708 [hep-ph]].
  
\bibitem{Armesto:2009fj} 
  N.~Armesto, L.~Cunqueiro and C.~A.~Salgado,
  Eur.\ Phys.\ J.\ C {\bf 63}, 679 (2009)
  [arXiv:0907.1014 [hep-ph]].
  
\bibitem{Schenke:2009gb} 
  B.~Schenke, C.~Gale and S.~Jeon,
  Phys.\ Rev.\ C {\bf 80}, 054913 (2009)
  [arXiv:0909.2037 [hep-ph]].
  
\bibitem{Majumder:2010qh}
A.~Majumder, M.~Van Leeuwen,
  [arXiv:1002.2206 [hep-ph]].

\bibitem{CasalderreySolana:2010eh} 
  J.~Casalderrey-Solana, J.~G.~Milhano and U.~A.~Wiedemann,
  J.\ Phys.\ G {\bf 38}, 035006 (2011)
  [arXiv:1012.0745 [hep-ph]].



\bibitem{Wang:2013cia} 
  X.~N.~Wang and Y.~Zhu,
  Phys.\ Rev.\ Lett.\  {\bf 111}, no. 6, 062301 (2013)
  [arXiv:1302.5874 [hep-ph]].

\bibitem{Zapp:2013vla} 
  K.~C.~Zapp,
  Eur.\ Phys.\ J.\ C {\bf 74}, no. 2, 2762 (2014)
  [arXiv:1311.0048 [hep-ph]].

\bibitem{Ghiglieri:2015zma} 
  J.~Ghiglieri and D.~Teaney,
  Int.\ J.\ Mod.\ Phys.\ E {\bf 24}, no. 11, 1530013 (2015)
  [arXiv:1502.03730 [hep-ph]].
  

  
\bibitem{Blaizot:2015lma} 
  J.~P.~Blaizot and Y.~Mehtar-Tani,
  Int.\ J.\ Mod.\ Phys.\ E {\bf 24}, no. 11, 1530012 (2015)
  [arXiv:1503.05958 [hep-ph]].
  
\bibitem{Chien:2015hda} 
  Y.~T.~Chien and I.~Vitev,
  JHEP {\bf 1605}, 023 (2016)
  [arXiv:1509.07257 [hep-ph]].
   
     
   
   
\bibitem{Cao:2017zih} 
  S.~Cao {\it et al.} [JETSCAPE Collaboration],
  Phys.\ Rev.\ C {\bf 96}, no. 2, 024909 (2017)
  [arXiv:1705.00050 [nucl-th]].
  
\bibitem{Arleo:2017ntr} 
  F.~Arleo,
  Phys.\ Rev.\ Lett.\  {\bf 119}, no. 6, 062302 (2017)
  [arXiv:1703.10852 [hep-ph]].
  






\bibitem{Casalderrey-Solana:2014bpa} 
  J.~Casalderrey-Solana, D.~C.~Gulhan, J.~G.~Milhano, D.~Pablos and K.~Rajagopal,
  JHEP {\bf 1410}, 019 (2014)
  Erratum: [JHEP {\bf 1509}, 175 (2015)]
  [arXiv:1405.3864 [hep-ph]].
  
\bibitem{Casalderrey-Solana:2015vaa} 
  J.~Casalderrey-Solana, D.~C.~Gulhan, J.~G.~Milhano, D.~Pablos and K.~Rajagopal,
  JHEP {\bf 1603}, 053 (2016)
  [arXiv:1508.00815 [hep-ph]].

\bibitem{Hulcher:2017cpt} 
  Z.~Hulcher, D.~Pablos and K.~Rajagopal,
  JHEP {\bf 1803}, 010 (2018)
  [arXiv:1707.05245 [hep-ph]].
  
\bibitem{Casalderrey-Solana:2018wrw} 
  J.~Casalderrey-Solana, Z.~Hulcher, G.~Milhano, D.~Pablos and K.~Rajagopal,
  Phys.\ Rev.\ C {\bf 99}, no. 5, 051901 (2019)
  [arXiv:1808.07386 [hep-ph]].
  
\bibitem{Sjostrand:2014zea} 
  T.~Sj\" ostrand {\it et al.},
  Comput.\ Phys.\ Commun.\  {\bf 191}, 159 (2015)
  [arXiv:1410.3012 [hep-ph]].
  
  \bibitem{Eskola:2009uj} 
  K.~J.~Eskola, H.~Paukkunen and C.~A.~Salgado,
  JHEP {\bf 0904}, 065 (2009)
  [arXiv:0902.4154 [hep-ph]].
  
\bibitem{CasalderreySolana:2011gx} 
  J.~Casalderrey-Solana, J.~G.~Milhano and P.~Quiroga-Arias,
  Phys.\ Lett.\ B {\bf 710}, 175 (2012)
  [arXiv:1111.0310 [hep-ph]].
  
\bibitem{Shen:2014vra} 
  C.~Shen, Z.~Qiu, H.~Song, J.~Bernhard, S.~Bass and U.~Heinz,
  Comput.\ Phys.\ Commun.\  {\bf 199}, 61 (2016)
  [arXiv:1409.8164 [nucl-th]].
  
\bibitem{Chesler:2014jva} 
  P.~M.~Chesler and K.~Rajagopal,
  Phys.\ Rev.\ D {\bf 90}, no. 2, 025033 (2014)
  [arXiv:1402.6756 [hep-th]].
  
\bibitem{Chesler:2015nqz} 
  P.~M.~Chesler and K.~Rajagopal,
  JHEP {\bf 1605}, 098 (2016)
  [arXiv:1511.07567 [hep-th]].
  
\bibitem{MehtarTani:2010ma} 
  Y.~Mehtar-Tani, C.~A.~Salgado and K.~Tywoniuk,
  Phys.\ Rev.\ Lett.\  {\bf 106}, 122002 (2011)
  [arXiv:1009.2965 [hep-ph]].
  
\bibitem{MehtarTani:2011tz} 
  Y.~Mehtar-Tani, C.~A.~Salgado and K.~Tywoniuk,
  Phys.\ Lett.\ B {\bf 707}, 156 (2012)
  [arXiv:1102.4317 [hep-ph]].
  
\bibitem{CasalderreySolana:2011rz} 
  J.~Casalderrey-Solana and E.~Iancu,
  JHEP {\bf 1108}, 015 (2011)
  [arXiv:1105.1760 [hep-ph]].
  
\bibitem{CasalderreySolana:2012ef} 
  J.~Casalderrey-Solana, Y.~Mehtar-Tani, C.~A.~Salgado and K.~Tywoniuk,
  Phys.\ Lett.\ B {\bf 725}, 357 (2013)
  [arXiv:1210.7765 [hep-ph]].
  
\bibitem{He:2015pra} 
  Y.~He, T.~Luo, X.~N.~Wang and Y.~Zhu,
  Phys.\ Rev.\ C {\bf 91}, 054908 (2015)
  Erratum: [Phys.\ Rev.\ C {\bf 97}, no. 1, 019902 (2018)]
  [arXiv:1503.03313 [nucl-th]].
  
  
\bibitem{Chen:2017zte} 
  W.~Chen, S.~Cao, T.~Luo, L.~G.~Pang and X.~N.~Wang,
  Phys.\ Lett.\ B {\bf 777}, 86 (2018)
  [arXiv:1704.03648 [nucl-th]].
  
\bibitem{Tachibana:2017syd} 
  Y.~Tachibana, N.~B.~Chang and G.~Y.~Qin,
  Phys.\ Rev.\ C {\bf 95}, no. 4, 044909 (2017)
  [arXiv:1701.07951 [nucl-th]].
 
\bibitem{He:2018xjv} 
  Y.~He, S.~Cao, W.~Chen, T.~Luo, L.~G.~Pang and X.~N.~Wang,
  Phys.\ Rev.\ C {\bf 99}, no. 5, 054911 (2019)
  [arXiv:1809.02525 [nucl-th]].
  
\bibitem{Park:2018acg} 
  C.~Park, S.~Jeon and C.~Gale,
  Nucl.\ Phys.\ A {\bf 982}, 643 (2019)
  [arXiv:1807.06550 [nucl-th]].
  
\bibitem{Chang:2019sae} 
  N.~B.~Chang, Y.~Tachibana and G.~Y.~Qin,
  arXiv:1906.09562 [nucl-th].
  
  
\bibitem{Mehtar-Tani:2017web} 
  Y.~Mehtar-Tani and K.~Tywoniuk,
  Phys.\ Rev.\ D {\bf 98}, no. 5, 051501 (2018)
  [arXiv:1707.07361 [hep-ph]].
  
\bibitem{Acharya:2017goa} 
  S.~Acharya {\it et al.} [ALICE Collaboration],
  Phys.\ Lett.\ B {\bf 776}, 249 (2018)
  [arXiv:1702.00804 [nucl-ex]].
  
\bibitem{Cacciari:2008gp} 
  M.~Cacciari, G.~P.~Salam and G.~Soyez,
  JHEP {\bf 0804}, 063 (2008)
  [arXiv:0802.1189 [hep-ph]].

  
\bibitem{Chatrchyan:2013kwa} 
  S.~Chatrchyan {\it et al.} [CMS Collaboration],
  Phys.\ Lett.\ B {\bf 730}, 243 (2014)
  [arXiv:1310.0878 [nucl-ex]].
  
\bibitem{Sirunyan:2017bsd} 
  A.~M.~Sirunyan {\it et al.} [CMS Collaboration],
  Phys.\ Rev.\ Lett.\  {\bf 120}, no. 14, 142302 (2018)
  [arXiv:1708.09429 [nucl-ex]].
   
\bibitem{Sirunyan:2018gct} 
  A.~M.~Sirunyan {\it et al.} [CMS Collaboration],
  JHEP {\bf 1810}, 161 (2018)
  [arXiv:1805.05145 [hep-ex]].
  
\bibitem{Acharya:2019djg} 
  S.~Acharya {\it et al.} [ALICE Collaboration],
  arXiv:1905.02512 [nucl-ex].
  
\bibitem{Kauder:2017cvz} 
  K.~Kauder [STAR Collaboration],
  Nucl.\ Part.\ Phys.\ Proc.\  {\bf 289-290}, 137 (2017)
  [arXiv:1703.10933 [nucl-ex]].
  
\bibitem{Dokshitzer:1997in} 
  Y.~L.~Dokshitzer, G.~D.~Leder, S.~Moretti and B.~R.~Webber,
  JHEP {\bf 9708}, 001 (1997)
  [hep-ph/9707323].
  
\bibitem{Wobisch:1998wt} 
  M.~Wobisch and T.~Wengler,
  In *Hamburg 1998/1999, Monte Carlo generators for HERA physics* 270-279
  [hep-ph/9907280].
  
\bibitem{Dasgupta:2013ihk}
  M.~Dasgupta, A.~Fregoso, S.~Marzani and G.~P.~Salam,
  JHEP {\bf 1309} (2013) 029
  [arXiv:1307.0007 [hep-ph]].
  
  
\bibitem{Larkoski:2015lea} 
  A.~J.~Larkoski, S.~Marzani and J.~Thaler,
  Phys.\ Rev.\ D {\bf 91}, no. 11, 111501 (2015)
  [arXiv:1502.01719 [hep-ph]].

\bibitem{Dreyer:2018nbf} 
  F.~A.~Dreyer, G.~P.~Salam and G.~Soyez,
  JHEP {\bf 1812}, 064 (2018)
  [arXiv:1807.04758 [hep-ph]].
  
\bibitem{Andrews:2018jcm} 
  H.~A.~Andrews {\it et al.},
  arXiv:1808.03689 [hep-ph].

\bibitem{Frye:2017yrw} 
  C.~Frye, A.~J.~Larkoski, J.~Thaler and K.~Zhou,
  JHEP {\bf 1709}, 083 (2017)
  [arXiv:1704.06266 [hep-ph]].

 
  
\bibitem{Chien:2016led} 
  Y.~T.~Chien and I.~Vitev,
  Phys.\ Rev.\ Lett.\  {\bf 119}, no. 11, 112301 (2017)
  [arXiv:1608.07283 [hep-ph]].

\bibitem{Chang:2017gkt} 
  N.~B.~Chang, S.~Cao and G.~Y.~Qin,
  Phys.\ Lett.\ B {\bf 781}, 423 (2018)
  [arXiv:1707.03767 [hep-ph]].
  
\bibitem{Mehtar-Tani:2016aco} 
  Y.~Mehtar-Tani and K.~Tywoniuk,
  JHEP {\bf 1704}, 125 (2017)
  [arXiv:1610.08930 [hep-ph]].
  
\bibitem{Blaizot:2013hx} 
  J.~P.~Blaizot, E.~Iancu and Y.~Mehtar-Tani,
  Phys.\ Rev.\ Lett.\  {\bf 111}, 052001 (2013)
  [arXiv:1301.6102 [hep-ph]].
  

  
\bibitem{Zapp:2012ak} 
  K.~C.~Zapp, F.~Krauss and U.~A.~Wiedemann,
  JHEP {\bf 1303}, 080 (2013)
  [arXiv:1212.1599 [hep-ph]].
  
\bibitem{KunnawalkamElayavalli:2017hxo} 
  R.~Kunnawalkam Elayavalli and K.~C.~Zapp,
  JHEP {\bf 1707}, 141 (2017)
  [arXiv:1707.01539 [hep-ph]].
  
\bibitem{Milhano:2017nzm} 
  G.~Milhano, U.~A.~Wiedemann and K.~C.~Zapp,
  Phys.\ Lett.\ B {\bf 779}, 409 (2018)
  [arXiv:1707.04142 [hep-ph]].

\bibitem{Caucal:2019uvr} 
  P.~Caucal, E.~Iancu and G.~Soyez,
  arXiv:1907.04866 [hep-ph].
  
  

\bibitem{Ellis:1991qj} 
  R.~K.~Ellis, W.~J.~Stirling and B.~R.~Webber,
  Camb.\ Monogr.\ Part.\ Phys.\ Nucl.\ Phys.\ Cosmol.\  {\bf 8}, 1 (1996).


\bibitem{Sjostrand:2007gs} 
  T.~Sj\" ostrand, S.~Mrenna and P.~Z.~Skands,
  Comput.\ Phys.\ Commun.\  {\bf 178}, 852 (2008)
  [arXiv:0710.3820 [hep-ph]].

  
\bibitem{Sjostrand:2004ef} 
  T.~Sj\" ostrand and P.~Z.~Skands,
  Eur.\ Phys.\ J.\ C {\bf 39}, 129 (2005)
  [hep-ph/0408302].

\end{thebibliography}
\end{document}